\documentclass[preprint,prb,aps,floats,amssymb,showpacs,draft,%
floatfix,superscriptaddress]{revtex4} 
\usepackage{psfig}

\def\singlespace{ \renewcommand{\baselinestretch}{1} \large\normalsize }

\begin{document}

\title{ 
The critical behavior of three-dimensional Ising spin glass models
} 

\singlespace

\author{Martin Hasenbusch} 
\affiliation{ 
Institut f\"ur Theoretische Physik, Universit\"at Leipzig, 
Postfach 100 920, D-04009 Leipzig, Germany.
} 
\author{Andrea Pelissetto} 
\affiliation{Dipartimento di Fisica
  dell'Universit\`a di Roma ``La Sapienza" and INFN, Roma, Italy.}
\author{Ettore Vicari} 
\affiliation{ 
Dipartimento di Fisica dell'Universit\`a di Pisa and INFN, Pisa, Italy.  } 

\date{\today}

\begin{abstract}
  We perform high-statistics Monte Carlo simulations of three-dimensional
  Ising spin-glass models on cubic lattices of size $L$: the $\pm J$
  (Edwards-Anderson) Ising model for two values of the disorder parameter $p$,
  $p=0.5$ and $p=0.7$ (up to $L=28$ and $L=20$, respectively), and the
  bond-diluted bimodal model for bond-occupation probability $p_b = 0.45$ (up
  to $L=16$). The finite-size behavior of the quartic cumulants at the
  critical point allows us to check very accurately that these models belong
  to the same universality class. Moreover, it allows us to estimate the
  scaling-correction exponent $\omega$ related to the leading irrelevant
  operator: $\omega=1.0(1)$.  Shorter Monte Carlo simulations of the
  bond-diluted bimodal models at $p_b=0.7$ and $p_b=0.35$ (up to $L=10$) and
  of the Ising spin-glass model with Gaussian bond distribution (up to $L=8$)
  also support the existence of a unique Ising spin-glass universality class.
  A careful finite-size analysis of the Monte Carlo data which takes into
  account the analytic and the nonanalytic corrections to scaling allows us to
  obtain precise and reliable estimates of the critical exponents $\nu$ and
  $\eta$: we obtain $\nu=2.45(15)$ and $\eta=-0.375(10)$.
\end{abstract}



\maketitle

\singlespace

\section{Introduction}
\label{intro}

The Ising model with random ferromagnetic and antiferromagnetic couplings 
is a simplified model\cite{EA-75} for
disordered uniaxial magnetic materials which 
show glassy behavior in some region of their phase
diagram, such as Fe$_{1-x}$Mn$_x$TiO$_3$ and Eu$_{1-x}$Ba$_x$MnO$_3$; see,
e.g., Refs.~\onlinecite{IATKST-86,GSNLAI-91,NN-07}.  The random nature of the
short-ranged interactions is mimicked by nearest-neighbor random bonds.
This model is  also interesting {\em per se}, since it provides 
a laboratory in which the combined effect of quenched disorder and 
frustration can be studied.  

It is now well established that three-dimensional Ising
spin-glass models present a high-temperature paramagnetic phase
and, for some values of the parameters, a low-temperature glassy phase 
(if the frustration is small, the low-temperature phase is 
ferromagnetic). The two phases 
are separated by a continuous phase
transition, which is expected to have universal features, 
i.e. to belong to a universality class which is independent of the
details of the model and, in particular, of the disorder distribution.  
Several numerical works 
\cite{OM-85,Ogielski-85,McMillan-85,BM-85,BY-85,SC-86,RZ-86,
  BY-88,MPR-94,KY-96,
  BPC-96,IPR-96,BJ-98,MPR-98,MC-99,PC-99,BCFMPRTTUU-00,MC-02,NEY-03,DCA-04,
  PC-05,PRT-06,PPV-06,Jorg-06,CHT-06,KKY-06,MNS-08,HPV-08,Belletti-etal-08}
have addressed these issues, considering various Ising spin-glass models,
characterized by different disorder distributions, with or without dilution.
Over the years many 
estimates of the critical exponents have been obtained.
We mention the most recent ones for the correlation-length exponent $\nu$: 
$\nu=2.39(5)$,\cite{KKY-06} $\nu=2.72(8)$,\cite{CHT-06}
$\nu=1.5(3)$,\cite{NEY-03} $\nu=1.35(10)$,\cite{MC-02}
$\nu=2.15(15)$,\cite{BCFMPRTTUU-00} $\nu=1.8(2)$,\cite{PC-99} obtained from
simulations of the symmetric model with bimodal distribution; $\nu=2.22(15)$
for the bond-diluted symmetric bimodal model with $p_b=0.45$;\cite{Jorg-06}
$\nu=2.44(9)$\cite{KKY-06} and $\nu=2.00(15)$~\cite{MPR-98} for the symmetric
model with Gaussian disorder distribution; $\nu=2.4(6)$ for the
random-anisotropy Heisenberg model in the limit of large
anisotropy,\cite{PPV-06} which is expected to be in the same Ising spin-glass
universality class.\cite{PPV-06,CL-77,LLMPV-07} Moreover, the analysis of
different quantities has often given different estimates of the same critical
exponent, even in the same model.  For instance, recent Monte Carlo (MC)
studies \cite{KKY-06,CHT-06} find significant discrepancies among the
estimates of the exponent $\nu$ obtained from the
finite-size scaling (FSS) at $T_c$ of the temperature derivative of $\xi/L$,
of the Binder cumulant, and of the overlap susceptibility. For the bimodal
Ising model Ref.~\onlinecite{KKY-06} quotes $\nu=2.39(5)$, $\nu=2.79(11)$, and
$\nu=1.527(8)$, from the analysis of these three quantities,
respectively.  
A likely reason for these discrepancies is the presence of scaling corrections,
which may be quite important in spin-glass systems since the absence of fast
MC algorithms makes it necessary to work with systems of relatively small
size.

One of the aims of the present paper is a detailed analysis of the role of
scaling corrections in spin-glass systems.  We present a general 
renormalization-group (RG) analysis
based on the Wegner expansion,\cite{Wegner-76} which allows us to predict the
corrections to the asymptotic critical behavior for the different quantities.
In particular, we show that the analytic dependence of the relevant scaling
fields on the model parameters, such as the temperature, may give rise to
scaling corrections that decay as powers of $L^{-1/\nu}$, 
where $L$ is the linear size of the lattice. Since $\nu\approx
2.45$ in Ising spin-glass systems, they decay quite slowly and may give rise
to systematic deviations which are difficult to detect, given the small
interval of values of $L$ which can be considered in MC simulations.  Their
presence explains some inconsistencies in the standard analyses of MC data
reported in the literature.  Thus, it is crucial to take scaling corrections
into account for an accurate study of the critical behavior, for a
robust check of universality among different models, and for
reliable estimates of universal quantities such as the critical exponents.

In this paper we report a high-statistics MC study of different Ising
spin-glass models. We consider the $\pm J$ Ising model for two values of the
disorder parameter, the bond-diluted symmetric bimodal model with various
values of the dilution, and also the model with Gaussian disorder
distribution.  We determine the FSS behavior of several 
RG invariant quantities, such as the ratio $R_\xi\equiv\xi/L$ ($\xi$ is 
the second-moment correlation length) and the
quartic cumulants defined from the overlap variables. We verify with good
precision their independence on the model and disorder distribution, providing
an accurate evidence of universality.  Then, we obtain an estimate of the
leading correction-to-scaling exponent $\omega$: $\omega=1.0(1)$. Finally, we
determine the critical exponents. We analyze the MC data at the critical point
and in the high-temperature phase, taking into account the RG predictions for
the scaling corrections and the precise above-reported estimate of $\omega$.
We obtain
\begin{eqnarray}
&& \nu=2.45(15), \nonumber \\
&&  \eta=-0.375(10).  
\end{eqnarray}
Then,
using scaling and hyperscaling relations we obtain
\begin{eqnarray}
&& \beta=\nu(1+\eta)/2=0.77(5), \nonumber \\ 
&& \gamma=(2-\eta)\nu=5.8(4), \nonumber \\
&& \alpha=2-3\nu=-5.4(5). 
\end{eqnarray}
In this work we extend the results presented in Ref.~\onlinecite{HPV-08}.
First, we have significantly increased the statistics of the large-$L$ data
for the bimodal symmetric Ising model and we have added some data for other
diluted models and for the bimodal model with Gaussian distributed couplings.
Second, we present a much more detailed analysis of the critical-point data
and a new analysis of the high-temperature data obtained in the
parallel-tempering simulations.  This allows us to check the universality of
the critical behavior of the correlation length and of the susceptibility in
the high-temperature phase. Finally, we discuss the extended-scaling scheme,
\cite{CHT-06,CHT-07} which is inspired by the high-temperature expansion.  As
already noted in Ref.~\onlinecite{CHT-06}, this scheme shows an apparent
improvement of the scaling behavior with respect to the naive approach in
which scaling corrections are simply neglected, at least for some quantities,
e.g., the overlap susceptibility.  However, as we shall show, such an
improvement is only marginal for the purpose of obtaining accurate results.
Indeed, this requires to take into account the analytic and nonanalytic
scaling corrections predicted by the RG theory.

The paper is organized as follows. In Sec.~\ref{secmod} we define the models
we investigate and the quantities that are computed in the MC simulation. In
Sec.~\ref{fsssec} we derive the FSS predictions of the RG theory which are the
basis of our FSS analyses. Some details are reported in App.~\ref{proofsca}.
In Sec.~\ref{mcsim} and in App.~\ref{MCinfo} we give some details on the MC
simulations.  In Sec.~\ref{unisevi} we discuss universality, verifying that
the infinite-volume limit of the quartic cumulants and of $R_\xi \equiv \xi/L$
is independent of the model. In Sec.~\ref{omegasec} we compute the leading
correction-to-scaling exponent $\omega$. In Sec.~\ref{crittemp} we compute the
critical exponents $\nu$ and $\eta$ and the critical temperature for the
different models by using the data close to the critical point.  In
Sec.~\ref{HT-est} we present a global analysis of all available
high-temperature data obtained in our parallel-tempering MC simulations. We
again determine the critical exponents and show that the FSS behavior of
several quantities is universal. Moreover, we discuss the extended-scaling
scheme of Ref.~\onlinecite{CHT-06}.  In Sec.~\ref{HTTL} we compute the
high-temperature zero-momentum quartic couplings.  Finally, in
Sec.~\ref{conclusions} we present our conclusions.

\section{Ising spin glass systems}
\label{secmod}

\subsection{The $\pm J$ Edwards-Anderson Ising model and its phase diagram}
\label{modsec1}

We consider the $\pm J$ Edwards-Anderson Ising model on a simple cubic
lattice of linear size $L$ with periodic boundary conditions. The
corresponding Hamiltonian is~\cite{EA-75}
\begin{equation}
H = - \sum_{\langle xy \rangle} J_{xy} \sigma_x \sigma_y,
\label{lH}
\end{equation}
where $\sigma_x=\pm 1$, the sum is over the nearest-neighbor lattice sites,
and the exchange interactions $J_{xy}$ are uncorrelated quenched random
variables with probability distribution
\begin{equation}
P(J_{xy}) = p \delta(J_{xy} - 1) + (1-p) \delta(J_{xy} + 1).  
\label{pmjdi}
\end{equation}
The usual bimodal Ising spin glass model, for which $[J_{xy}]=0$
(brackets indicate the average over the disorder distribution), corresponds to
$p=1/2$.  For $p\neq 1/2$ we have
\begin{equation}
[J_{xy}]=2p-1\neq 0,
\end{equation}
and ferromagnetic (or antiferromagnetic) configurations are energetically
favored. Note that 
the free energy and also the correlations of the overlap variables 
that we shall define below are invariant under $p\to 1-p$
and thus we can always assume $p \ge 1/2$. 

\begin{figure*}[tb]
\centerline{\psfig{width=9truecm,angle=0,file=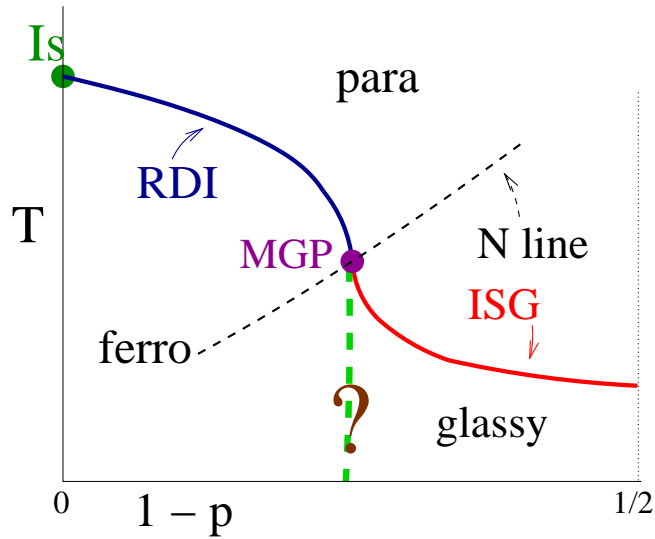}}
\caption{
  Phase diagram of the three-dimensional $\pm J$ 
  (Edwards-Anderson) Ising model in the $T$-$p$
  plane for $p \ge 1/2$, i.e., $1-p \le 1/2$. 
  The phase diagram is symmetric under $p\rightarrow 1-p$.
  }
\label{phdiad3}
\end{figure*}

The $T$-$p$ phase diagram of the three-dimensional $\pm J$ Ising model is
sketched in Fig.~\ref{phdiad3}.
The high-temperature phase is paramagnetic for any $p$.
The nature of the low-temperature phase depends on the value of $p$: it is
ferromagnetic for small values of $1-p$, while it is glassy with vanishing
magnetization for sufficiently large values of $1-p$.  The three
phases are separated by 
transition lines, which meet at a magnetic-glassy multicritical point (MGP),
usually called Nishimori point, which is located along the so-called Nishimori
line~\cite{Nishimori-81,GHDB-85,LH-88,Nishimori-book,HPPV-07-mgp}
defined by the relation ($p \ge 1/2$)
\begin{equation}
T=T_N(p),\qquad T_N(p) = {2\over \ln {p\over 1 - p}}\,.
\label{Nishimori-eq}
\end{equation}
On the Nishimori line the magnetic and the overlap two-point correlation 
functions are equal.  

The paramagnetic-ferromagnetic (PF) transition line starts at the Ising
transition at $p=1$ and extends up to the MGP at $p=p^*$.  For $p=1$ there is
the standard Ising transition at~\cite{DB-03} $T_{\rm Is}=4.5115248(14)$.
Disorder is a relevant perturbation of the pure three-dimensional Ising fixed
point. As a consequence, the Ising point $p=1$ is a multicritical point
\cite{HPPV-07-pmj} with crossover exponent $\phi=\alpha_{\rm Is}$,
where\cite{CPRV-02} $\alpha_{\rm Is}=0.1096(5)$ is the Ising specific-heat
exponent.  The critical behavior for any $1>p>p^*$ belongs to the
randomly-dilute Ising (RDI) universality class,\cite{HPPV-07-pmj}
characterized by the correlation-length critical exponent \cite{HPPV-07,PV-02}
$\nu=0.683(2)$ and by the magnetic exponent $\eta=0.036(1)$. 
The coordinates of the MGP in the $T$-$p$
plane are~\cite{HPPV-07-mgp} $T^*= 1.6692(3)$ and $p^*=0.76820(4)$.  The
multicritical behavior is characterized by~\cite{HPPV-07-mgp} the thermal
exponent $\nu=1.64(5)$, the crossover exponent $\phi=1.67(10)$, and 
the magnetic (and also overlap) exponent 
$\eta = -0.114(3)$.

The paramagnetic-glassy (PG) transition line starts at the MGP and extends up
to $p=1/2$ (actually up to $p=1-p^*=0.23180(4)$ due to the symmetry $p\to 1-p$
of the phase diagram).  A reasonable hypothesis is that the PG critical
behavior is independent of $p$, as found in mean-field
models.\cite{Toulouse-80} Hence, for any $p^*> p > 1-p^*$ the PG transition is
expected to belong to the same universality class (named ISG in
Fig.~\ref{phdiad3}) as that of the bimodal Ising spin-glass model at $p=1/2$.
The critical behavior along this transition line is the main subject of this
paper.  As we shall see, the universality hypothesis is fully confirmed by our
FSS analyses at $p=0.5$ and $p=0.7$.

The nature of the ferromagnetic-glassy (FG) transition line is not clear yet.
At fixed $p$ the following inequality holds:\cite{Nishimori-81,KR-03}
\begin{equation}
| [ \langle \sigma_x \sigma_y \rangle_T ] | \le
[ | \langle \sigma_x \sigma_y \rangle_{T_N(p)} | ] ,
\label{ineq}
\end{equation}
where the subscripts indicate the temperature of the thermal average, and
$T_N(p)$ is the temperature along the Nishimori line, defined in
Eq.~(\ref{Nishimori-eq}).  This relation shows that ferromagnetism can only
exist in the region $p>p^*$ and that the system is maximally magnetized along
the Nishimori line.  Ref.~\onlinecite{Nishimori-86} (see also
Refs.~\onlinecite{Kitatani-92,Nishimori-book}) also argues that the FG
transition line coincides with the line $p=p^*$, from $T=T^*$ to $T=0$.
Recent numerical investigations~\cite{WHP-03,AH-04,PHP-06} of the
two-dimensional $\pm J$ model have shown that this conjecture is not exact,
though quantitative deviations are small: at $T=0$ the critical value $p_c$
where ferromagnetism disappears is definitely larger than $p^*$, indicating a
reentrant transition line.  In three dimensions Ref.~\onlinecite{Hartmann-99}
quotes $p_c=0.778(5)$, which is slightly larger than $p^* = 0.76820(4)$, with
an associated critical exponent $\nu=1.1(3)$: it is therefore likely that the
conjecture does not hold in three dimensions as well.  We also mention that a
mixed low-temperature phase,\cite{Kitatani-94} in which ferromagnetism and
glassy order coexist, is found in mean-field models~\cite{Toulouse-80} such as
the infinite-range Sherrington-Kirkpatrick model.\cite{SK-75} Its presence has
been confirmed in the $\pm J$ Ising model on a Bethe lattice.\cite{CKR-05}
However, there is no evidence of a mixed phase in the $\pm J$ Ising model on a
cubic lattice~\cite{Hartmann-99} and in related models.\cite{KM-02} In
particular, the numerical results of Ref.~\onlinecite{Hartmann-99} show that
the onset of the glassy behavior at $T=0$ occurs close to the point where the
ferromagnetic phase disappears, and are consistent with a single transition
within numerical precision.  Nevertheless, the existence of such a mixed phase
is still an open problem, as discussed in Ref.~\onlinecite{CKR-05}.

\subsection{Other Ising spin-glass models}
\label{modsec2}

We also consider the bond-diluted bimodal Ising model (BDBIM) defined by 
Hamiltonian (\ref{lH}) with bond probability distribution
\begin{equation}
P(J_{xy}) = p_b \, \left[ {1\over 2} \delta(J_{xy} - J) + 
{1\over 2} \delta(J_{xy} + J)\right] + 
(1-p_b) \delta(J_{xy}) ~.
\label{probdisd}
\end{equation}
A PG transition occurs for sufficiently small
values of $1-p_b$, i.e. for $p_b > p_{SG}$. While investigations 
at $T=0$ indicate that $p_{SG}$ should be identified with the 
bond-percolation
point~\cite{BM-08,BD-08} ($p_{\rm perc}=0.2488126(5)$ on a simple cubic
lattice\cite{LZ-98}), recent investigations of the critical behavior close to 
the percolation point suggest that $p_{SG}$ is larger than
$p_{\rm perc}$.\cite{JR-08}

The model can be extended by considering the distribution
\begin{equation}
P(J_{xy}) = p_b \, \left[ p \delta(J_{xy} - J) + (1-p)\delta(J_{xy} + J)\right] + 
(1-p_b) \delta(J_{xy}) .
\label{probdisde}
\end{equation}
In this case, for $p_b>p_{SG}$ ($p_{SG}$ may depend on $p$) 
we expect a $T$-$p$ phase diagram
similar to the one sketched in Fig.~\ref{phdiad3} for the $\pm J$ Ising
model without dilution, with a PF 
and a PG transition line meeting at a multicritical point.

A PG transition is also observed in the
random-bond Ising spin-glass model with Gaussian bond distribution:
\begin{equation}
P(J_{xy}) = {1\over \sqrt{2\pi}} e^{-J_{xy}^2/2}.
\label{gaussdist}
\end{equation}
This transition is expected to be in the same universality class
as that of the bimodal Ising model.\cite{KKY-06}

\subsection{Overlap thermodynamic quantities}
\label{overth}

In this work we focus on the critical behavior of the overlap parameter
\begin{equation}
q_x \equiv  \sigma_x^{(1)} \sigma_x^{(2)}, 
\label{ovpar}
\end{equation}
where the spins $\sigma_x^{(i)}$ belong to two independent replicas with the
same disorder realization $\{J_{xy}\}$. The corresponding correlation function
is
\begin{equation}
G(x) \equiv [ \langle q_0 q_x \rangle ] = [ \langle \sigma_0\sigma_x\rangle^2 ],
\label{twop}
\end{equation}
where the angular and square brackets indicate the
thermal average and the quenched average over $\{J_{xy}\}$, respectively. 
We define the
susceptibility $\chi$ and the second-moment correlation length $\xi$ as 
\begin{eqnarray}
&&\chi \equiv  \sum_{x} G(x), \label{chidef} \\
&&\xi^2 \equiv  {1\over 4 \sin^2 (p_{\rm min}/2)} 
{\widetilde{G}(0) - \widetilde{G}(p)\over \widetilde{G}(p)},
\label{xidefffxy}
\end{eqnarray}
where $p = (p_{\rm min},0,0)$, $p_{\rm min} \equiv 2 \pi/L$, and
$\widetilde{G}(q)$ is the Fourier transform of $G(x)$.

We also define the RG invariant quantities
\begin{eqnarray}
&&R_\xi\equiv\xi/L,\label{rxidef}\\
&&U_{4}  \equiv { [ \mu_4 ]\over [\mu_2]^{2}}, \label{cum4}\\
&&U_{22} \equiv  {[ \mu_2^2 ]-[\mu_2]^2 \over [\mu_2]^2},
\label{cum22}
\end{eqnarray}
where 
\begin{equation}
\mu_{k} \equiv \langle \; ( \sum_x q_x\; )^k \rangle. 
\end{equation}
We call them phenomenological couplings and denote them by $R$ in the
following.

In the high-temperature paramagnetic phase, we also consider the zero-momentum
quartic couplings
\begin{eqnarray}
&&G_4 \equiv - \frac{\chi_4}{\xi^3 \chi^2},
\label{g4def}
\\
&&G_{22} \equiv - \frac{\chi_{22}}{\xi^3 \chi^2},
\label{g22def}
\end{eqnarray}
where
\begin{eqnarray}
&&\chi_4 \equiv \frac{1}{L^3} \left ([\mu_4] -3 [\mu_2^2] \right),
\label{chi4def}
\\
 &&\chi_{22} \equiv \frac{1}{L^3} \left ([\mu_2^2] -[\mu_2]^2 \right).
\label{chi22def}
\end{eqnarray}
The critical limit $T\to T_c^+$ of the zero-momentum quartic couplings $G_4$
and $G_{22}$ is universal.

\section{Finite-size scaling}
\label{fsssec}

In this section we summarize some basic results concerning FSS, which allow us
to understand the role of the {\em analytic} and {\em nonanalytic} scaling
corrections.  We consider two Ising spin-glass systems coupled by 
an interaction 
\begin{equation}
h \sum_x q_x = h \sum_x \sigma^{(1)}_x \sigma^{(2)}_x,
\end{equation}
where $h$ is a constant external field. The model is defined on a 
cubic lattice of linear size $L$ with periodic boundary conditions.

By applying standard RG
arguments we expect the disorder-averaged free-energy density 
to be the sum of
a regular part and a singular part:
\begin{equation}
{\cal F}(\beta,h,L) = 
{\cal F}_{\rm reg}(\beta,h,L) + {\cal F}_{\rm sing}(\beta,h,L),
\label{FscalL0}
\end{equation}
where $\beta\equiv 1/T$.
The regular part is expected to depend on $L$ only through exponentially small
terms, while the singular part encodes the critical behavior.  The starting
point of FSS is the scaling behavior of the singular part of the free-energy
density  (see, e.g., Refs.~\onlinecite{Wegner-76,Privman-90,SS-00,PV-02}):
\begin{eqnarray}
{\cal F}_{\rm sing}(\beta,h,L) =
 L^{-d} F( u_h L^{y_h}, u_t L^{y_t}, \{ v_i L^{y_i}\}),
\label{FscalL01}
\end{eqnarray}
where $d$ is the space dimension, 
$u_h$ and $u_t$ are the scaling fields associated with $h$
and the reduced temperature $t\sim 1-\beta/\beta_c$ (their RG dimensions are
$y_h=(d+2-\eta)/2$ and $y_t=1/\nu$, respectively),
and $v_i$ are irrelevant scaling fields
with $y_i<0$.  At the critical point we have $u_t(t=0,h=0) = 0$ and
$u_h(t=0,h=0) = 0$, while, generically, we expect $v_i(t=0,h=0)\not=0$.  Since
$y_i<0$, for large $L$ 
the free energy can be expanded in powers of $\{ v_i L^{y_i}\}$.
Therefore, we can write
\begin{eqnarray}
{\cal F}_{\rm sing}(\beta,h,L) 
= L^{-d} f(u_h L^{y_h}, u_t L^{y_t}) +
  v_\omega L^{-d-\omega} f_{\omega}(u_h L^{y_h}, u_t L^{y_t})  + \ldots
\label{FscalL}
\end{eqnarray}
where the leading {\em nonanalytic} correction-to-scaling exponent $\omega$ is
related to the RG dimension $y_\omega$ of the leading irrelevant scaling field
$v_\omega\equiv v_1$, $\omega= - y_\omega$.
The scaling fields are analytic
functions of the system parameters---in particular, of $h$ and $t$---and are
expected not to depend on $L$. Note also that the size $L$ is expected to be
an exact scaling field for periodic boundary conditions. For a general
discussion of these issues, see Ref.~\onlinecite{SS-00}, Sec.~III of 
Ref.~\onlinecite{CHPV-06}, and references
therein.  In general, $u_t$ and $u_h$ can be expanded as
\begin{eqnarray}
&&u_h = h \bar{u}_h(t) + O(h^3),\quad \bar{u}_h(t) = a_h + a_1 t + O(t^2),
\label{uhsca}\\
&&u_t = c_t t + c_{02} t^2 + c_{20} h^2 + c_{21} h^2 t + O(t^3,h^4,h^4 t)  ,
\label{utsca}
\end{eqnarray}
where we used the fact that the free energy is symmetric under $h\to -h$. In
the expansion of $u_{h,t}$ around the critical point $h,t=0$, the terms beyond
the leading ones give rise to {\em analytic} scaling corrections. 

The scaling behavior of zero-momentum thermodynamic quantities can be obtained
by performing appropriate derivatives of ${\cal F}$ with respect to
$h$.  For instance, for the overlap susceptibility at $h = 0$ we obtain
\begin{equation}
\chi(\beta,L) = \left. { \partial^2 {\cal F}\over \partial h^2}\right |_{h=0} =
L^{2-\eta} \bar{u}_h(t)^2 g(u_t L^{y_t}) \left[
1 + v_\omega L^{-\omega} g_\omega(u_t L^{y_t}) + \ldots \right] + 
g_{\rm reg}(\beta).
\label{chiexp_1}
\end{equation}
The function $g_{\rm reg}(\beta)$ represents the contribution of the regular
part ${\cal F}_{\rm reg}(\beta,h,L)$ of the free-energy density and is $L$
independent (apart from exponentially small terms). It
gives rise to a correction proportional to $L^{\eta -2}$.
Analogous formulae hold for the $2n$-point
susceptibilities.

The FSS of the phenomenological couplings is given by
\begin{eqnarray}
R(\beta,L) &=&  r(u_t L^{y_t}) + v_\omega r_\omega(u_t L^{y_t}) L^{-\omega}  + 
\ldots
\nonumber \\
&=& 
R^* + r'(0) c_t \, t L^{y_t}   + \ldots +  c_\omega \, L^{-\omega} + \ldots,
\label{Rexp_1}
\end{eqnarray}
where $R^*\equiv r(0)$, $c_\omega = v_\omega r_\omega(0)$,
and the second line holds only very close to the 
critical point, for $t L^{y_t} \ll 1$.  A proof of
Eq.~(\ref{Rexp_1}) for the phenomenological couplings $U_4$, $U_{22}$, and
$R_\xi$ is presented in App.~\ref{proofsca}.

The thermal RG exponent $y_t=1/\nu$ is usually computed from the FSS of the
derivative $R'$ of a phenomenological coupling $R$ with respect to $\beta$ at
$\beta_c$.  Using Eq.\ (\ref{Rexp_1}) one obtains
\begin{eqnarray}
R' \equiv {\partial R\over \partial\beta} =  
L^{y_t} (\partial_\beta{u}_t) \left[ r'(u_t L^{y_t}) +
v_\omega L^{-\omega } r_\omega'(u_t L^{y_t}) + \cdots \right] .
\label{scaling-derivative-R}
\end{eqnarray}
One may also consider the derivative $\chi'\equiv d\chi/d\beta$ of the 
susceptibility $\chi$.
From Eq.~(\ref{chiexp_1})
we obtain
\begin{eqnarray}
\chi' &=& 
L^{2-\eta+y_t} \bar{u}_h^2 \partial_\beta{u}_t 
\left\{ g'(u_t L^{y_t}) + 
    v_\omega L^{-\omega} \left[
     g'(u_t L^{y_t}) g_\omega(u_t L^{y_t}) + 
     g(u_t L^{y_t}) g'_\omega(u_t L^{y_t}) \right]\right\}
\nonumber \\
&& + 2 L^{2-\eta} \bar{u}_h \partial_\beta\bar{u}_h g(u_t L^{y_t}) + \cdots
+ g'_{\rm reg}(\beta).
\label{chidexp_1}
\end{eqnarray}
Note that the second term in the right hand side  gives rise to
scaling corrections proportional to $L^{-y_t}= L^{-1/\nu}$, while the
background term $g'_{\rm reg}(\beta)$ leads to corrections proportional to 
$L^{-y_t-2+\eta}$.

At $T=T_c$, setting $t=0$ in the above-reported equations, we obtain:
\begin{eqnarray}
&& R = R^* + c_\omega L^{-\omega} + \ldots, \label{rin}\\
&& \chi = c L^{2-\eta} ( 1 + c_\omega L^{-\omega} + \ldots ), 
\label{chibc} \\
&& R' = c L^{1/\nu} ( 1  + c_\omega L^{-\omega} + \ldots), \label{Rpbc} \\
&& \chi' =  c L^{2-\eta+1/\nu} 
( 1  + c_\omega L^{-\omega} + \ldots + c_a L^{-1/\nu} + \ldots ) .
\label{chipbc}
\end{eqnarray}
Note that, unlike the temperature derivative $R'$ of a RG-invariant quantity,
$\chi'$ also presents an $L^{-1/\nu}$ scaling correction, due to the analytic
dependence on $t$ of the scaling field $u_h$ (for this reason we call it {\em
analytic} correction).  Since, as we shall see, in the 
Ising spin-glass case $1/\nu\approx 0.4$ and
$\omega\approx 1.0$, the scaling corrections in $\chi'$ decay 
significantly more slowly than
those occurring in $R'$. This makes the ratio
\begin{equation}
{\chi'\over \chi} \sim L^{1/\nu}
\label{ratiochi}
\end{equation}
unsuitable for a precise determination of $\nu$ and explains the significant
discrepancies observed in Ref.~\onlinecite{KKY-06}.

Instead of computing the various quantities at fixed Hamiltonian parameters,
one can also consider FSS keeping a phenomenological coupling $R$ fixed at 
a given
value $R_{f}$.\cite{Has-99,HPPV-07}  This means that, for each $L$, one
determines $\beta_f(L,R_f)$, 
such that $R(\beta=\beta_f(L,R_f),L) = R_{f}$, 
and then considers any quantity at $\beta = \beta_f(L,R_f)$. 
The value $R_{f}$ can be
specified at will, as long as $R_f$ is taken between the high- and
low-temperature fixed-point values of $R$.  For $R_f\not=R^*$,
where $R^*$ is defined in Eq.~(\ref{rin}),
$\beta_f$ converges to $\beta_c$ as 
\begin{equation}
\beta_f-\beta_c \sim L^{-1/\nu}, 
\label{betafco1}
\end{equation}
while for
$R_{f} = R^*$ we have
\begin{equation}
\beta_f-\beta_c \sim L^{-1/\nu-\omega}.
\label{betafco2}
\end{equation}
Indeed, if $u_{t,f}(L,R_f)$ is the value of $u_t$ for $\beta = \beta_f(L,R_f)$,
we obtain from Eq.~(\ref{Rexp_1})
\begin{equation}
u_{t,f} L^{y_t} = B(R_f) + v_\omega B_\omega(R_f) L^{-\omega} + \cdots
\label{Expan-ut}
\end{equation}
where, using Eq.~(\ref{Rexp_1}),
\begin{eqnarray}
r(B(x)) &=& x \; , \\
B_\omega(x) &=& - {r_\omega(B(x))\over r'(B(x))}\; .
\end{eqnarray}
Now, if $R_f = R^*$, we have $B(R_f) = 0$, which implies 
$u_{t,f} \sim L^{-y_t-\omega}$, hence Eq.~(\ref{betafco2}).
Otherwise, $B(R_f)$ is different from zero and we obtain the behavior 
(\ref{betafco1}).

If we now substitute relation (\ref{Expan-ut}) into 
Eqs.~(\ref{chiexp_1}), (\ref{Rexp_1}), (\ref{scaling-derivative-R}), and 
(\ref{chidexp_1}), we obtain the expansion of the different quantities at
fixed $R_f$, which we denote by adding a bar: given ${\cal O}(\beta,L)$,
we define $\bar{\cal O}(L,R_f) \equiv {\cal O}[\beta_f(L,R_f),L]$.
For $R_f = R^*$, since $u_{t,f} \sim L^{-y_t - \omega}$ we reobtain 
Eqs.~(\ref{rin}), (\ref{chibc}), (\ref{Rpbc}), and (\ref{chipbc}),
with different coefficients, of course. 
If $R_f\not = R^*$, we must be more careful. 
The behavior of another
phenomenological coupling $R_\alpha$ does not change 
qualitatively and we still have
\begin{equation}
\bar{R}_\alpha(L,R_f) \approx
\bar{R}_\alpha^* + \overline{c}_\alpha L^{-\omega} + \ldots ,
\label{barr}
\end{equation}
where $\bar{R}_\alpha^* = r_\alpha[B(R_f)]$ is universal. 
It depends on $R_f$ and satisfies $\bar{R}_\alpha^* = R^*_\alpha$
for $R_f = R^*$. Also $\bar{\chi}'$ behaves as it does at fixed $T=T_c$, i.e. 
it follows Eq.~(\ref{chipbc}).
On the other hand, 
$\bar{\chi}$ and $\bar{R}'$ present additional analytic corrections.
Indeed, since [see Eq.~(\ref{uhsca})]
\begin{equation}
\bar{u}_{h,f} = a_h + {a_1\over c_t} B(R_f) L^{-y_t} + \cdots
\end{equation}
(nonanalytic $O(L^{-\omega})$ corrections have been neglected),
Eq.~(\ref{chiexp_1}) gives
\begin{equation}
\bar{\chi}_\alpha(L,R_f) = 
   L^{2-\eta} \left[ a_h^2 + 2 {a_1 a_h \over c_t} B(R_f) L^{-y_t} + 
    O(L^{-2y_t})\right] g(B(R_f)) [1 + O(L^{-\omega})] .
\end{equation}
If $R_f\not=R^*$, $B(R_f)\not=0$, and thus analytic corrections
occur. Note that, if $R_f$ is close to $R^*$, since $B(R^*) = 0$, 
we have 
\begin{equation}
B(R_f)\sim R_f - R^*.
\end{equation}
Hence, in this case the analytic corrections are small, of order 
$(R_f - R^*) L^{-1/\nu}$. In general, corrections of 
order $L^{-k/\nu}$ have amplitudes proportional to
$(R_f - R^*)^k$.

\section{Monte Carlo simulations}
\label{mcsim}

In the MC simulations we employed the Metropolis algorithm, the random-exchange
method (often called parallel-tempering or 
multiple Markov-chain method),\cite{raex} and multispin coding.
See App.~\ref{MCinfo} for details on their implementation.

We simulated the $\pm J$ Ising model at $p=0.5$ for $L=3$-14,16,20,24,28, at
$p=0.7$ for $L=3$-12,14,16,20, and the BDBIM at $p_b=0.45$ for
$L=4$-12,14,16.  We averaged over a large number $N_s$ of disorder samples:
$N_s\approx 6.4\cdot 10^6$ up to $L=12$, $N_s/10^3\approx
2400,2800,1500,245,150,18$, respectively for $L=13,14,16,20,24,28$ in the case
of the $\pm J$ Ising model at $p=0.5$.  Similar statistics were collected at
$p=0.7$ (except for $L=20$ where the statistics were approximately 1/3
of those for $p=0.5$), while for the
BDBIM statistics were smaller (typically, by a factor of two for
the small lattices and by a factor of 6 for the largest ones).
We also considered the BDBIM at $p_b=0.7$ and $p_b=0.35$ and the 
Ising model with Gaussian distributed couplings, but only performed simulations
for small values of $L$:
$L=4,6,8,10$ for the BDBIM and $L=4,5,6,8$ for the Gaussian model.

For each $L$ and model we performed parallel-tempering runs.  This allowed us
to estimate the different quantities in a large interval $[\beta_{\rm
  min},\beta_{\rm max}]$. To fix $\beta_{\rm max}$ we used the results of
Ref.~\onlinecite{KKY-06}, which provided the best estimates of $R_\xi^*$ at
the time we started our simulations: 0.627(4) and 0.635(9) for an Ising model
with bimodal and Gaussian distributed bonds, respectively.  Thus, in most of
the runs $\beta_{\rm max}$ was chosen so that $R_\xi(\beta_{\rm max},L)\approx
0.63$.  Only in the most recent simulations (two runs with $L=20$ and 24) of
the $\pm J$ Ising model at $p=0.5$ did we take $\beta_{\rm max}$ such that
$R_\xi(\beta_{\rm max},L)\approx 0.66$.  We checked thermalization by using
the recipe outlined in Ref.~\onlinecite{KKY-06}, see App.~\ref{MCinfo} for
some details.

As already emphasized in Refs.~\onlinecite{BFMMPR-98-b,HPPV-07}, in
high-precision MC studies of random systems one should be careful when
computing disorder averages of products of thermal expectations, for instance
the quartic cumulant $U_{22}$ defined in Eq.~(\ref{cum22}). Indeed, naive
estimators have a bias that may become significantly larger than the
statistical error if $N_s$ is large. We use (essentially) bias-free
estimators, defined as discussed in Ref.~\onlinecite{HPPV-07}; some details
are given in App.~\ref{MCinfo}.

In total, the MC simulations took approximately 
40 years of CPU time on a single core
of a 2.4 GHz AMD Opteron processor.

\section{Universality and correction-to-scaling exponent $\omega$}
\label{quarticcum}

\subsection{Analysis of the quartic cumulants at fixed $R_\xi$}
\label{unisevi}

\begin{table}
\squeezetable
\caption{Quartic cumulant
$\bar{U}_4$ at fixed $R_\xi=0.63$ for 
the $\pm J$ Ising model at $p=0.5$ and $p=0.7$,
for the BDBIM at $p_b=0.45,\,0.7,\,0.35$,
and for the Ising spin-glass model with Gaussian bond distribution.
}
\label{u4data}
\begin{ruledtabular}
\begin{tabular}{rllllll}
\multicolumn{1}{c}{$L$}&
\multicolumn{1}{c}{$\pm J_{p=0.5}$}&
\multicolumn{1}{c}{$\pm J_{p=0.7}$}&
\multicolumn{1}{c}{BDBIM$_{p_b=0.45}$}&
\multicolumn{1}{c}{BDBIM$_{p_b=0.7}$}&
\multicolumn{1}{c}{BDBIM$_{p_b=0.35}$}&
\multicolumn{1}{c}{Gaussian}\\
\colrule
4 & 1.48231(6)& 1.46813(5)& 1.49036(8)& 1.48480(6)& 1.49164(9)& 1.49145(13) \\
5 & 1.48985(6) & 1.47597(6) & 1.49853(8) &  &  & 1.4996(2) \\
6 & 1.49446(6) & 1.48193(6) & 1.50300(9) & 1.49618(9) & 1.50788(9) & 1.5033(2) \\
7 & 1.49753(6) & 1.48642(6) & 1.50544(9) &  &  &  \\
8 & 1.49987(6) & 1.48984(6) & 1.50714(9) & 1.50082(9) & 1.51320(13)& 1.5063(5)\\
9 & 1.50136(6) & 1.49260(6) & 1.50815(9) &  &  & \\
10& 1.50273(6) & 1.49478(6) & 1.50889(9) & 1.50382(11)& 1.5146(2)  & \\
11& 1.50383(6) & 1.49665(6) & 1.50946(9) &&&\\
12& 1.50469(6) & 1.49781(7) & 1.50984(13)&&&\\
13& 1.50541(11)&            &            &&&\\
14& 1.50618(10)& 1.50030(12)& 1.5103(3)  &&&\\
16& 1.50702(13)& 1.50220(13)& 1.5113(3)  &&&\\
20& 1.5081(3)  & 1.5048(5)  &            &&&\\
24& 1.5089(4)  &  &  &&&\\
28& 1.5108(13) &  &  &&&\\
\end{tabular}
\end{ruledtabular}
\end{table}

In this section we investigate the universality of the critical behavior of
Ising spin-glass models by comparing the 
limit for $L\rightarrow\infty$ of $U_4$ and $U_{22}$
computed at fixed $R_\xi$, denoted by $\bar{U}_4$ and $\bar{U}_{22}$,
respectively.  As discussed in Sec.~\ref{fsssec}, for sufficiently large $L$,
$\bar{U}_4$ and $\bar{U}_{22}$ are expected to behave as
\begin{equation}
\bar{U}_\# = \bar{U}_\#^* + c_\# L^{-\omega} ,
\label{barrl}
\end{equation}
where the constants $\bar{U}_\#^*$ are universal (model independent) but
depend on the fixed value of $R_\xi$; $\omega$ is the leading
scaling-correction exponent.

\begin{table}
\squeezetable
\caption{Quartic cumulant
  $\bar{U}_{22}$ at fixed $R_\xi=0.63$ for 
the $\pm J$ Ising model at $p=0.5$ and $p=0.7$,
for the BDBIM at $p_b=0.45,\,0.7,\,0.35$,
and for the Ising spin-glass model with Gaussian bond distribution.
}
\label{u22data}
\begin{ruledtabular}
\begin{tabular}{rllllll}
\multicolumn{1}{c}{$L$}&
\multicolumn{1}{c}{$\pm J_{p=0.5}$}&
\multicolumn{1}{c}{$\pm J_{p=0.7}$}&
\multicolumn{1}{c}{BDBIM$_{p_b=0.45}$}&
\multicolumn{1}{c}{BDBIM$_{p_b=0.7}$}&
\multicolumn{1}{c}{BDBIM$_{p_b=0.35}$}&
\multicolumn{1}{c}{Gaussian}\\
\colrule
4 & 0.13714(6) & 0.13955(6) & 0.13581(9) & 0.13602(7) & 0.14635(10)&0.13522(15)\\
5 & 0.14088(7) & 0.14087(6) & 0.13935(10)& & & 0.14014(24)\\
6 & 0.14277(7) & 0.14181(7) & 0.14166(10)& 0.14193(10)& 0.14501(10)& 0.14227(24)\\
7 & 0.14392(6) & 0.14262(7) & 0.14308(10)& & & \\
8 & 0.14478(7) & 0.14318(7) & 0.14415(10)& 0.14414(10)& 0.14597(15)& 0.1434(5)\\
9 & 0.14522(7) & 0.14380(7) & 0.14470(10)& & & \\
10& 0.14561(7) & 0.14418(7) & 0.14518(10)& 0.14536(12)& 0.14586(24)&\\
11& 0.14595(7) & 0.14453(7) & 0.14566(10)& & &\\
12& 0.14618(7) & 0.14465(7) & 0.14605(14)& & &\\
13& 0.14650(11)& & & & &\\
14& 0.14671(10)& 0.14531(13)& 0.1462(3) & & &\\
16& 0.14675(14)& 0.14553(14)& 0.1469(4) & & &\\
20& 0.1469(4)  & 0.1458(5)  & & & &\\
24& 0.1477(5)  & & & & &\\
28& 0.1496(14) & & & & &\\
\end{tabular}
\end{ruledtabular}
\end{table}

In Tables~\ref{u4data} and \ref{u22data} we report the estimates of
$\bar{U}_4$ and $\bar{U}_{22}$ at fixed $R_\xi=0.63$ for different models.
Without performing any analysis, one can immediately observe that the results
obtained for the different models are very close, and appear to approach the
same large-$L$ limit as $L$ increases. For instance, the estimates of
$\bar{U}_4$ for the largest lattices differ at most by 0.5\%, while those of
$\bar{U}_{22}$ vary by a few percent.  This already provides strong support to
universality. 

\begin{figure}[tb]
\centerline{\psfig{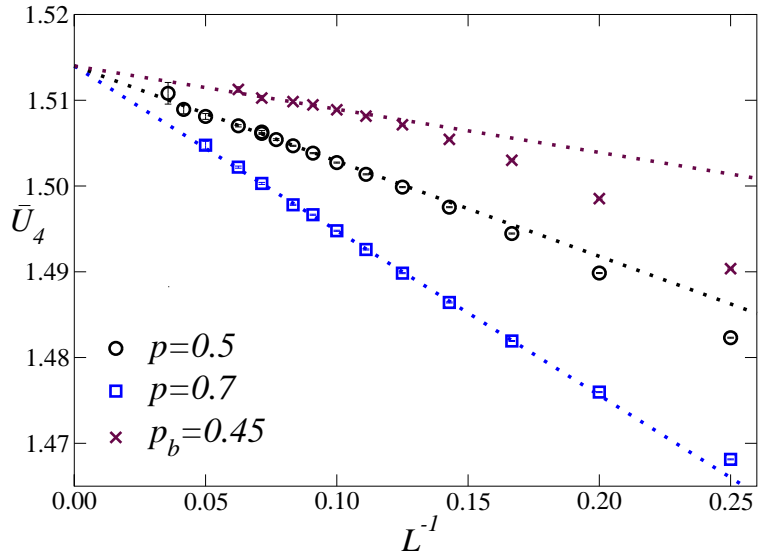}}
\vspace{2mm}
\caption{
  (Color online) Estimates of $\bar{U}_4(L)$ vs $L^{-1}$, for the $\pm J$ model at
  $p=0.5$ and $p=0.7$, and the BDBIM at $p_b=0.45$.  The dotted lines are
  drawn to guide the eye.  }
\label{u4}
\end{figure}

\begin{figure}[tb]
\centerline{\psfig{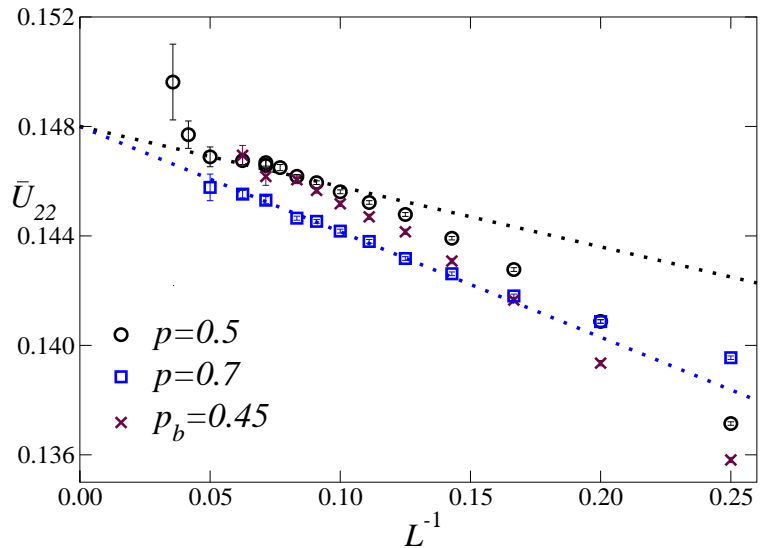}}
\vspace{2mm}
\caption{
  (Color online) Estimates of $\bar{U}_{22}(L)$ vs $L^{-1}$, 
  for the $\pm J$ model
  at $p=0.5$ and $p=0.7$, and the BDBIM  at $p_b=0.45$.  The dotted lines are
  drawn to guide the eye.  }
\label{u22}
\end{figure}

For a more detailed analysis,
let us first consider the three models for which we have most data, the
$\pm J$ model at $p=0.5$ and $p=0.7$, and the BDBIM at $p_b=0.45$. The MC
estimates of $\bar{U}_4(L)$ are shown in Fig.~\ref{u4} versus $1/L$.  The
results for the $\pm J$ Ising model at $p=0.5$ and $p=0.7$ fall quite nicely
on two straight lines that approach the same point as $L\to\infty$.  They
support the universality of the critical behavior and show the presence of 
scaling
corrections with an exponent $\omega\approx 1.0$.  In the case of the BDBIM with
$p_b=0.45$, the data apparently 
show a faster approach to the same infinite-volume limit.  
A fit of $\bar{U}_4(L)$ to $\bar{U}^*_4 + c L^{-\varepsilon}$ 
using all data with $L>4$ gives an effective exponent $\varepsilon\approx 2$.
However, 
for sufficiently large $L$, $L\gtrsim 10$ say, scaling corrections
are consistent with a linear $1/L$ behavior, as in the other models.
The corresponding amplitude $|c_4|$ is quite small, at least 
a factor of two smaller than in the undiluted bimodal model. 
This means that for $L\lesssim 10$ next-to-leading 
corrections to scaling dominate and determine the apparent approach
of the data to the infinite-volume limit.  Since the ratios of the
amplitudes of the leading scaling corrections are universal, a small $|c_4|$
implies that the leading nonanalytic scaling correction is small for any
quantity.  Thus, in this model the approach of any thermodynamic quantity to
the critical limit should be faster than in the $\pm J$ models
without dilution, as already noted in Ref.~\onlinecite{Jorg-06}, at least for
sufficiently large lattice sizes, where next-to-leading corrections
can be neglected.  Similar results are obtained for
$\bar{U}_{22}$, see Fig.~\ref{u22}.  However, in this case next-to-leading
corrections appear to be more important, since the $1/L$ behavior 
is observed for somewhat larger values of $L$.

\begin{table}
\squeezetable
\caption{
  Results of the fits of $\bar{U}_4$ and $\bar{U}_{22}$ at 
  fixed $R_\xi=0.63$ to $\bar{U}^* + c L^{-\varepsilon}$
  for several values of $L_{\rm min}$, the minimum lattice size 
  allowed in the fits.
  DOF is the number of degrees of freedom of the fit.
  The fits labelled ``$p=0.5$ \& $p=0.7$" are fits in 
  which the data for $p=0.5$ and $p=0.7$ are analyzed together, 
  assuming the universality of $\bar{U}_4^*$ and $\bar{U}_{22}^*$.
}
\label{tabu4u22}
\begin{ruledtabular}
\begin{tabular}{llrlrclr}
\multicolumn{1}{c}{Model}&
\multicolumn{1}{c}{}&
\multicolumn{1}{c}{$L_{\rm min}$}&
\multicolumn{1}{c}{$\bar{U}_4^*$}&
\multicolumn{1}{c}{$\chi^2/{\rm DOF}$}&
\multicolumn{1}{c}{$\quad$}&
\multicolumn{1}{c}{$\bar{U}_{22}^*$}&
\multicolumn{1}{c}{$\chi^2/{\rm DOF}$}\\
\colrule
$\pm J$,$p=0.5$ &free $\varepsilon$ &  6 & 1.5127(4)  & 0.7 && 0.1478(3) & 0.7 \\
        &                   &  8 & 1.5135(9)  & 0.8 && 0.1487(9) & 0.7 \\
        &                   & 10 & 1.5119(11) & 0.4 && 0.1482(12)& 0.8 \\
        & $\varepsilon=1$   &  8 & 1.5144(2)  & 0.8 && 0.1490(2) & 0.6 \\
        &                   & 12 & 1.5139(4)  & 0.6 && 0.1488(4) & 0.9 \\
\hline
$\pm J$,$p=0.7$ &free $\varepsilon$ &  6 & 1.5153(7) & 2.0 && 0.1481(8) & 0.8 \\
        &                   &  8 & 1.5145(14) & 2.7 && 0.1471(9) & 0.9 \\
        & $\varepsilon=1$   &  8 & 1.5143(2)  & 2.3 && 0.1478(2) & 0.9 \\
        &                   & 12 & 1.5153(5)  & 0.1 && 0.1482(5) & 0.8 \\
\hline
BDBIM,$p_b=0.45$ 
        &free $\varepsilon$ &  6 & 1.5120(3)  & 0.5 && 0.1479(6)  & 0.6 \\
        &                   &  9 & 1.5132(22) & 0.6 && 0.1485(23) & 0.5 \\
        & $\varepsilon=1$   &  8 & 1.5153(3)  & 0.8 && 0.1496(3) & 0.4 \\
        &                   & 12 & 1.5148(12)  & 1.2 && 0.1489(13) & 1.0 \\
\hline
$\pm J$,$p=0.5$ \& $p=0.7$
        & $\varepsilon=1$   &  8 & 1.5143(1)  & 1.2 && 0.1485(1) & 2.1 \\
        &                   & 12 & 1.5145(3)  & 0.9 && 0.1486(3) & 0.8 \\
        &                   & 16 & 1.5133(9)  & 0.4 && 0.1489(10)& 1.1 \\
        & $\varepsilon=1.2$ &  8 & 1.5120(1)  & 4.9 && 0.1479(1) & 2.5 \\
        &                   & 12 & 1.5127(3)  & 0.4 && 0.1482(3) & 0.9 \\
        &                   & 16 & 1.5123(8)  & 0.3 && 0.1486(9) & 1.1 \\
\end{tabular}
\end{ruledtabular}
\end{table}

\begin{figure*}[tb]
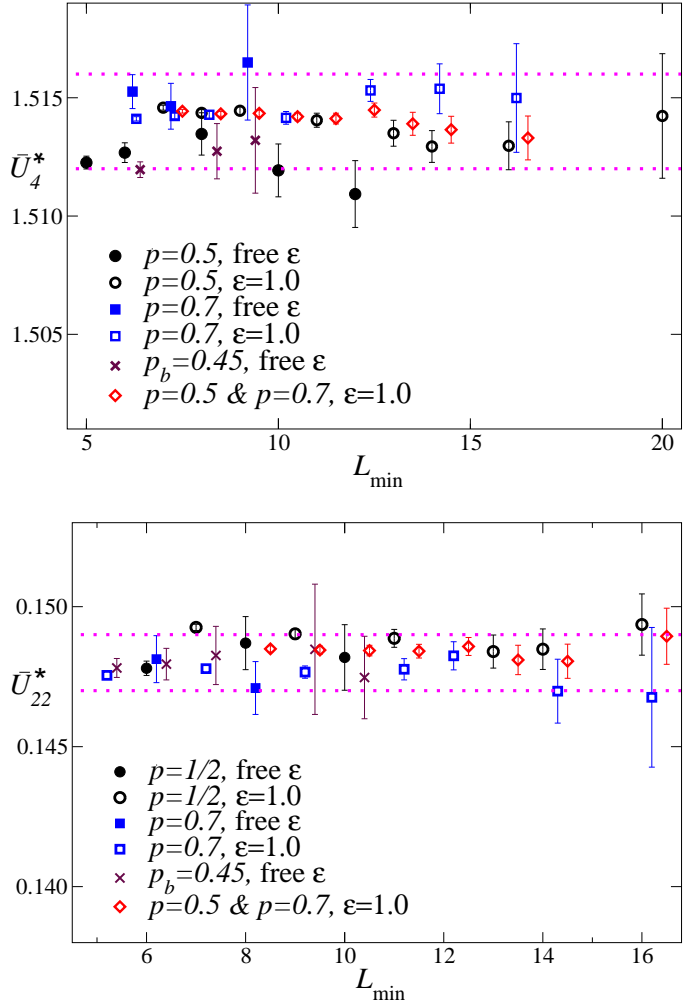

\centerline{\psfig{width=9truecm,angle=0,file=fig4a.eps}}
\vspace{4mm}
\centerline{\psfig{width=9truecm,angle=0,file=fig4b.eps}}
\vspace{2mm}
\caption{Estimates of $\bar{U}_4^*$ and $\bar{U}_{22}^*$, 
  obtained by fitting the quartic cumulants $\bar{U}_\#$ at fixed 
  $R_\xi=0.63$ to $\bar{U}^*_\# + c L^{-\varepsilon}$. 
  Here $L_{\rm min}$ is the minimum lattice size 
  allowed in the fits.
  The fits labelled ``$p=0.5$ \& $p=0.7$" are fits in 
  which the data for $p=0.5$ and $p=0.7$ are analyzed together.
  The dotted lines correspond to the final estimates
  $\bar{U}_4^*=1.514(2)$ and $\bar{U}_{22}^*=0.148(1)$.  }
\label{u4u22fit}
\end{figure*}

In order to estimate the universal infinite-volume values $\bar{U}_4^*$ and
$\bar{U}_{22}^*$, we perform fits of $\bar{U}_\#$ to $\bar{U}_\#^* + c_\#
L^{-\varepsilon}$. The exponent $\varepsilon$ is either a free parameter, 
or is set equal to 1,  which corresponds,
as we discuss below, to our best
estimate of the leading scaling-correction exponent $\omega$.  In
Table~\ref{tabu4u22} we report the results.  They are all consistent with the
estimates
\begin{equation}
\bar{U}_4^*=1.514(2), \qquad \bar{U}_{22}^*=0.148(1).  
\label{u4u22est}
\end{equation}
The accuracy and stability of the fits can also be appreciated in
Fig.~\ref{u4u22fit}, where the fit results are plotted versus the minimum
lattice size $L_{\rm min}$ allowed in the fits.  We can thus conclude that the
estimates of the quartic cumulants for the $\pm J$ Ising model and the BDBIM
at $p_b=0.45$ are fully consistent with universality. The relative differences
are approximately one per mille in the case of $\bar{U}_4$ and one per cent
for $\bar{U}_{22}$.

\begin{figure*}[tb]
\centerline{\psfig{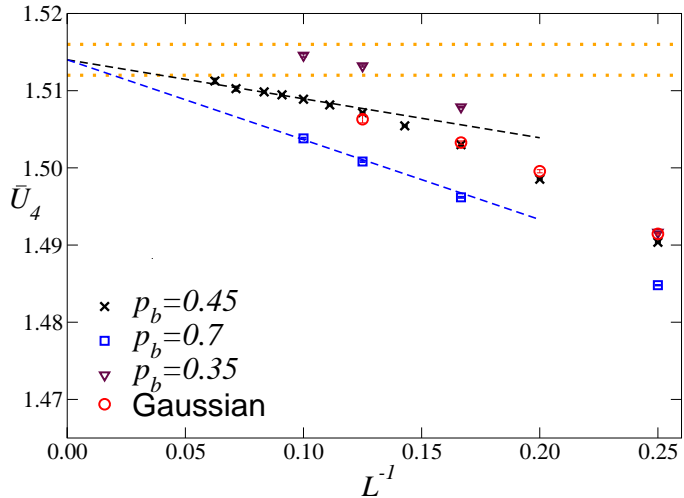}}
\vspace{2mm}
\caption{Estimates of $\bar{U}_4(L)$ for the BDBIM for several 
  values of $p_b$ and for the Ising spin-glass model with Gaussian bond
  distribution.  The dotted lines correspond to the estimate
  (\protect\ref{u4u22est}), $\bar{U}_4^*=1.514(2)$.  The dashed straight lines
  associated with the data at $p_b=0.7$ and $p_b=0.45$ show the linear
  behavior of the data for their largest lattices.
}
\label{u4BDBIMgen}
\end{figure*}

We also considered the BDBIM for the other dilution values, i.e., for $p_b=0.7$
and $p_b=0.35$, though in this case we have data only up to $L=10$.
The estimates of $\bar{U}_4$ are shown versus $1/L$ in Fig.~\ref{u4BDBIMgen}. 
For $p_b=0.7$ scaling corrections clearly behave as $1/L$
and are larger than at $p_b=0.45$. For the scaling-correction amplitude $c_4$
defined in Eq.~(\ref{barrl}) we obtain the estimate
$c_4\approx -0.10$, to be compared with the result $c_4\approx -0.11$
for the $\pm J$ model at $p=0.5$ and $c_4\approx -0.05$ 
for the diluted model at $p_b=0.45$. For $p_b=0.35$ 
we do not have enough data to estimate $c_4$; however, 
the data reported in Fig.~\ref{u4BDBIMgen} apparently 
approach the infinite-volume limit faster than at $p_b = 0.45$.
As indicated by the MC data
for $p_b\lesssim 0.27$ of Ref.~\onlinecite{JR-08},
the scaling corrections should increase as $p_b$ further decreases. 
This suggests that the
optimal model---the one with the smallest leading scaling
corrections---corresponds to a dilution in the range $0.3\lesssim p_b^*
\lesssim 0.4$: the model with $p_b=0.35$ should be close to the optimal one.

\begin{figure*}[tb]
\centerline{\psfig{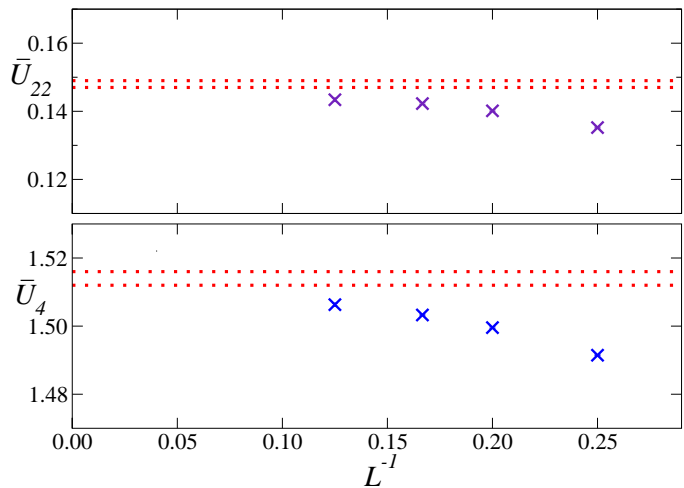}}
\vspace{2mm}
\caption{$\bar{U}_4$ (below) and $\bar{U}_{22}$ (above) for 
the Ising spin-glass model with Gaussian bond distribution.
The dotted lines correspond to the estimates (\protect\ref{u4u22est}),
$\bar{U}_4^*=1.514(2)$ and $\bar{U}_{22}^*=0.148(1)$.  }
\label{u4u22gau}
\end{figure*}

The estimates of $\bar{U}_4$ and $\bar{U}_{22}$ for the Ising spin-glass model
with Gaussian bond distribution are shown in Figs.~\ref{u4BDBIMgen} and
\ref{u4u22gau}: they are very close, and clearly approach, the estimates
(\ref{u4u22est}).  The results for $L=8$ differ by 0.5\% ($\bar{U}_4$) and 3\%
($\bar{U}_{22}$) from the asymptotic value.

In conclusion, the results for the $\pm J$ model, for the BDBIM, and for the
model with Gaussian distributed couplings
provide a very strong evidence of universality for the critical behavior of
Ising spin-glass models along the PG transition line.

\subsection{The RG exponent of the leading irrelevant operator}
\label{omegasec}

\begin{figure}[tb]
\centerline{\psfig{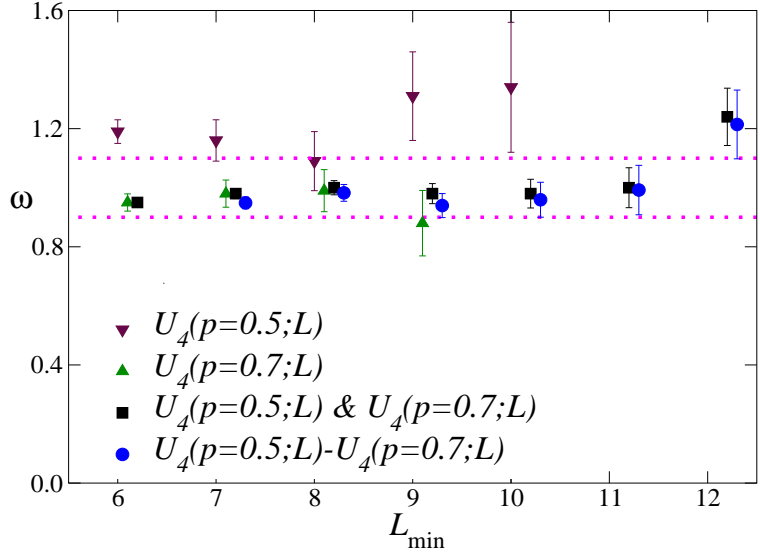}}
\vspace{2mm}
\caption{
  (Color online) Estimates of the leading correction-to-scaling exponent
  $\omega$ versus the minimum lattice size $L_{\rm min}$ used in
  the fit.  
  The fits labelled ``$p=0.5$ \& $p=0.7$" are fits in 
  which the data for $p=0.5$ and $p=0.7$ are analyzed together.
  The dotted lines correspond to the final estimate $\omega=1.0(1)$.
}
\label{omega}
\end{figure}

The analyses of $\bar{U}_4$ and $\bar{U}_{22}$ also give estimates of
$\omega$. The most precise ones are obtained from fits of $\bar{U}_4$.
In Fig.~\ref{omega} we show the estimates of $\omega$ 
obtained from fits of $\bar{U}_4$ to $\bar{U}^*_4 + c_p L^{-\omega}$ and 
from fits of the difference $\bar{U}_4(p=0.5;L)-\bar{U}_4(p=0.7;L)$ to
$bL^{-\omega}$. We estimate
\begin{equation}
\omega=1.0(1).
\label{omegaest}
\end{equation}
Consistent but less precise results are obtained from fits of 
$\bar{U}_{22}$.  The result (\ref{omegaest}) 
is consistent with 
the less precise estimates $\omega=0.84^{+0.43}_{-0.37}$ and
$\omega=1.0(4)$ reported in Refs.~\onlinecite{BCFMPRTTUU-00} 
and \onlinecite{PPV-06}, respectively.

In order to verify that scaling corrections are controlled by a 
single correction-to-scaling term with exponent $\omega = 1$ 
and we have not determined an effective exponent
arising from several almost degenerate exponents,
we check that the ratio $c_{22}/c_4$
is universal, where $c_\#$ is the scaling-correction amplitude appearing in
Eq.~(\ref{barrl}), as predicted by standard RG arguments.  In order to
determine this ratio, we fit together the available estimates of $\bar{U}(L)$
for the $\pm J$ Ising models at $p=0.5$ and $p=0.7$ to $\bar{U}(L) = \bar{U}^*
+ c_p L^{-\varepsilon}$, fixing $\varepsilon=1$ and taking $\bar{U}^*$ and the
two amplitudes $c_{p=0.5}$ and $c_{p=0.7}$ as free parameters.  Fits of the
data for $L\ge L_{\rm min}=10$ give
\begin{equation}
\bar{U}_4^*=1.5142(2),\quad c_4(p=0.5)=-0.114(2),\quad c_4(p=0.7)=-0.194(2),
\label{ampu4}
\end{equation}
($\chi^2/{\rm DOF}\approx 1.3$, where DOF is the number of degrees of
freedom of the fit), and
\begin{equation}
\bar{U}_{22}^*=0.1484(2),\quad c_{22}(p=0.5)=-0.027(2),\quad 
c_{22}(p=0.7)=-0.044(2),
\label{ampu22}
\end{equation}
($\chi^2/{\rm DOF}\approx 1.4$). It follows
\begin{eqnarray}
&& c_{22}/c_4=0.24(2) \quad {\rm for} \quad p=0.5,\\
&& c_{22}/c_4=0.23(1) \quad {\rm for} \quad p=0.7,
\end{eqnarray}
which are in good agreement.  These results are quite stable with increasing
$L_{\rm min}$. For $L_{\rm min}=12$ we obtain $c_{22}/c_4=0.24(3)$ and
$c_{22}/c_4=0.24(2)$ for $p=0.5$ and $p=0.7$, while for $L_{\rm min}=14$ we
find $c_{22}/c_4=0.19(9)$ and $c_{22}/c_4=0.21(5)$, respectively; in both
cases the fits have $\chi^2/{\rm DOF}\lesssim 1$.  Moreover, the variation of
the ratio $c_{22}/c_4$ with respect to a change of the exponent $\varepsilon$
by $\pm 0.1$ [it corresponds to the uncertainty of $\omega$, $\omega=1.0(1)$],
is smaller than the above-reported errors.  These results fully support our
interpretation of the $O(L^{-1.0})$ corrections as the corrections arising
from the leading irrelevant scaling field.

\section{Critical temperature and exponents}
\label{crittemp}

\subsection{Estimates of $\beta_c$}
\label{crittemp1}

\begin{figure}[tb]
\centerline{\psfig{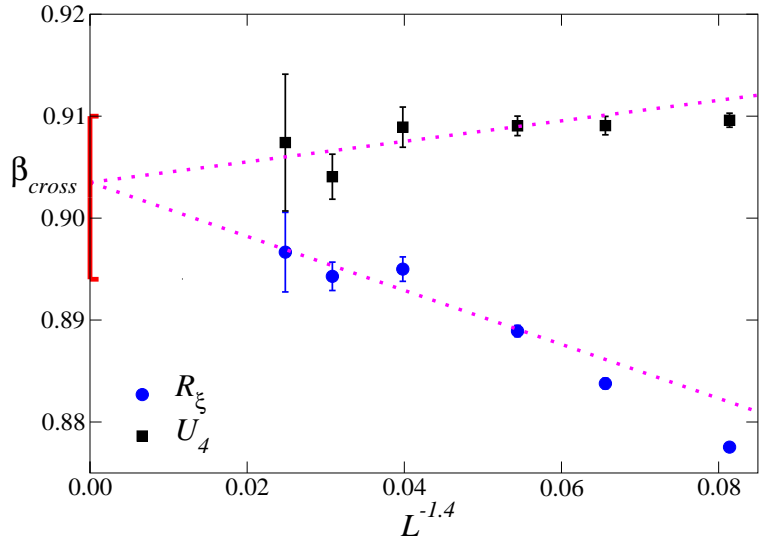}}
\vspace{2mm}
\caption{ (Color online) Crossing point 
  $\beta_{\rm cross}(L)$ for the
  $\pm J$ Ising model at $p=0.5$, from several pairs $L, 2L$, as
  obtained from $R_\xi$ and $U_4$, versus $L^{-1/\nu-\omega}\sim
  L^{-1.4}$. The dotted lines show the result of a fit to $\beta_c+ cL^{-1.4}$,
  which takes into account only the data satisfying
  $L\ge L_{\rm min}=8$---it corresponds to 
  $L_{\rm min}^{-1.4} \approx 0.054$. 
  The bar along the $y$-axis corresponds to the final
  estimate $\beta_c=0.902(8)$.  }
\label{bcross}
\end{figure}

\begin{figure}[tb]
\centerline{\psfig{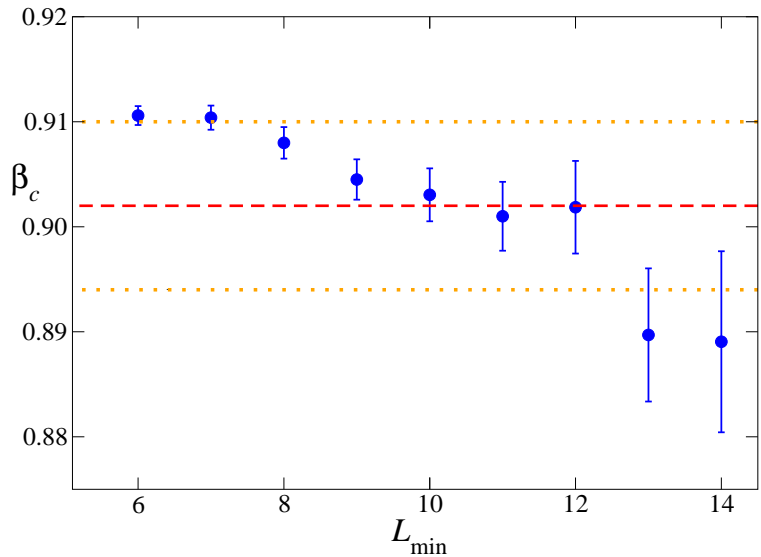}}
\vspace{2mm}
\caption{ (Color online) 
  Estimates of $\beta_c$ 
  as obtained from fits of $R_\xi$ and $U_4$ to Eq.~(\ref{rlb}) with
  $\omega=1$, for the $\pm J$ Ising model at $p=0.5$. All fits have 
  $\chi^2/{\rm DOF}<1.5$.  The lines correspond to the final
  estimate $\beta_c=0.902(8)$.  }
\label{bclmin}
\end{figure}

\begin{figure}[tb]
\centerline{\psfig{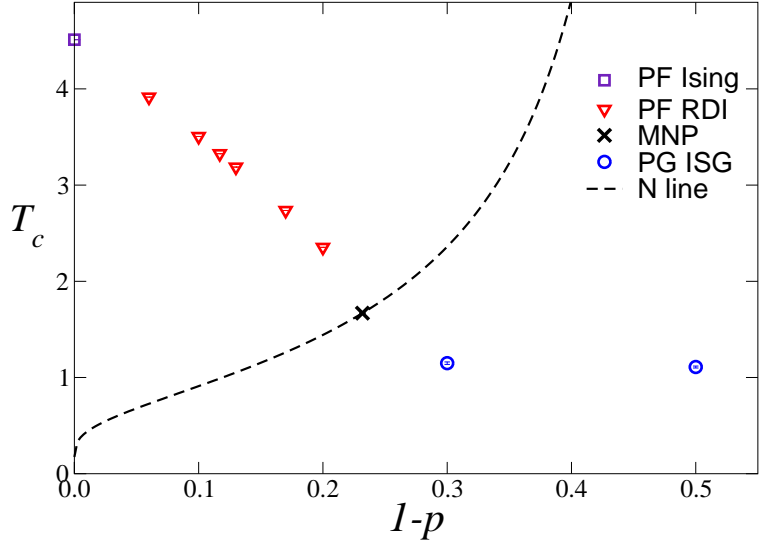}}
\vspace{2mm}
\caption{ (Color online) 
  The values of the critical temperature $T_c$ versus $1-p$ for the various
  transitions of the $\pm J$ Ising model.  The estimates at the
  paramagnetic-ferromagnetic transitions are 
  taken from Ref.~\onlinecite{DB-03} for
  the Ising transition, from Ref.~\onlinecite{HPPV-07-pmj} 
  for the RDI transition,
  and from Ref.~\onlinecite{HPPV-07-mgp} for that along the $N$ line.  }
\label{tcpmJ}
\end{figure}

In order to determine the critical temperature, we perform a standard FSS
analysis of $R_\xi$ and the quartic cumulants.
Figure~\ref{bcross} shows the crossing points $\beta_{\rm cross}(L)$ 
determined by solving the equation
\begin{equation}
R(\beta_{\rm cross},L) = R(\beta_{\rm cross},2L),
\label{betacross}
\end{equation}
for the $\pm J$ Ising model at $p=0.5$, using $R_\xi$ and $U_4$. 
In the large-$L$ limit 
$\beta_{\rm cross}(L)$ is expected to converge to $\beta_c$ as
\begin{equation}
\beta_{\rm cross}(L)-\beta_c \sim L^{-1/\nu-\omega} \sim L^{-1.4},
\label{convbcr}
\end{equation}
since $\nu\approx 2.45$ and $\omega\approx 1.0$. We perform a combined fit of
$R_\xi$ and $U_4$ to $\beta_c+ c_\# L^{-\varepsilon}$ with $\varepsilon=-1.4$,
taking $\beta_c$ and the two amplitudes $c_{R_\xi}$ and $c_{U_4}$ as free
parameters.  Using only the results with $L\ge L_{\rm min}=8$, we obtain
$\beta_c=0.9035(22)$ with $\chi^2/{\rm DOF}\approx 1.1$ (since the estimates
of $R_\xi$ and $U_4$ are correlated the error is only indicative). The
corresponding lines are drawn in Fig.~\ref{bcross}. We also consider
larger values of $L_{\rm min}$. We find that the estimates of $\beta_c$ vary
significantly (much more than the statistical error) when $L_{\rm min}$
varies, indicating that the systematic error due to the additional scaling
corrections is significantly larger than the statistical one.  Equivalently,
one can perform fits to
\begin{equation}
R(L,\beta_c) = R^* + b L^{-\omega},
\label{rlb}
\end{equation}
taking $R = R_\xi$ or $U_4$.\cite{foot-fit} The
results are shown in Fig.~\ref{bclmin}.  Taking into account the 
dependence of the results on $L_{\rm min}$ and their variation as 
$\omega$ varies by one error bar we obtain the final estimates
\begin{equation}
\beta_c=0.902(8),\qquad T_c=1.109(10).
\label{bcest}
\end{equation}
We regard the error as conservative.
We also consider the pseudocritical $\beta_f(L)$ defined in 
Sec.~\ref{fsssec} at fixed $R_\xi$, which converges to $\beta_c$
as $L\to\infty$, cf. Eqs.~(\ref{betafco1}) and
(\ref{betafco2}). The results are consistent with the estimate (\ref{bcest}).
The analysis of the crossing points and the fits to Eq.~(\ref{rlb}) 
also provide estimates 
of the limit $L\to\infty$ of $R_\xi$ and $U_4$ at the critical point.
We obtain
\begin{equation}
R_\xi^* = 0.645(15),\qquad U_4^*=1.50(2).
\label{tcest}
\end{equation}

Concerning the other models, similar analyses of $R_\xi$ and
$U_4$ give $\beta_c=0.87(1)$ [$T_c=1.149(13)$] for the $\pm J$ model at
$p=0.7$, and $\beta_c=1.54(2)$ [$T_c=0.649(8)$] for the BDBIM
at $p_b=0.45$.  The corresponding values of $R_\xi^*$ and $U_4^*$ are
consistent with the estimates (\ref{tcest}).

Finally, in Fig.~\ref{tcpmJ} we collect all results for the critical
temperature $T_c$ of the $\pm J$ Ising model at its various PF and PG
transitions as a function of the disorder parameter $p$.

\subsection{The critical exponent $\nu$}
\label{critexpnu}

\begin{figure}[tb]
\centerline{\psfig{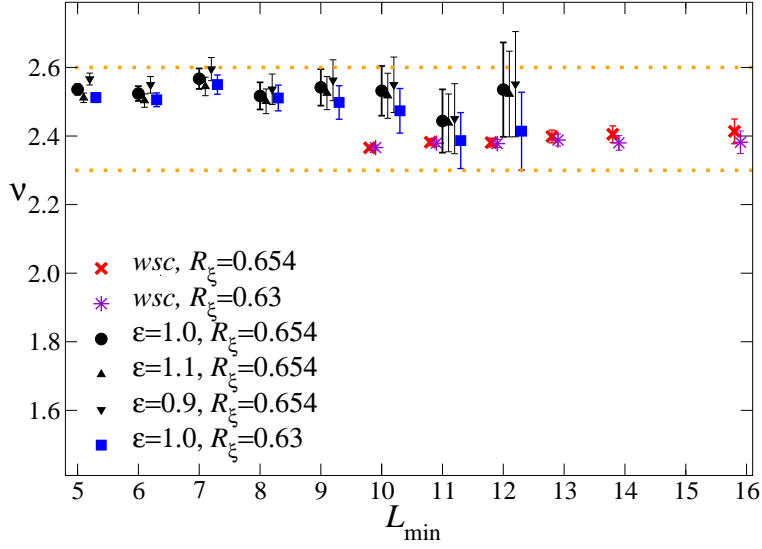}}
\vspace{2mm}
\caption{ (Color online) Estimates of the exponent $\nu$ from the analysis
  of $\bar{R}_\xi'$, at $R_\xi=0.63$ and $R_\xi=0.654$, for the 
  $\pm J$ Ising model
  at $p=0.5$, obtained by fitting the data without scaling corrections (wsc), 
  Eq.~(\ref{ansatznu0}), 
  and with scaling corrections, Eq.~(\ref{ansatznu}).  
  We only show the results of the fits which satisfy 
  $\chi^2/{\rm DOF}\lesssim 1.5$,
up to values of $L_{\rm min}$ where the errors blow up.
The dotted lines correspond to the final estimate $\nu=2.45(15)$.  }
\label{nup0p5rxi}
\end{figure}

\begin{figure}[tb]
\centerline{\psfig{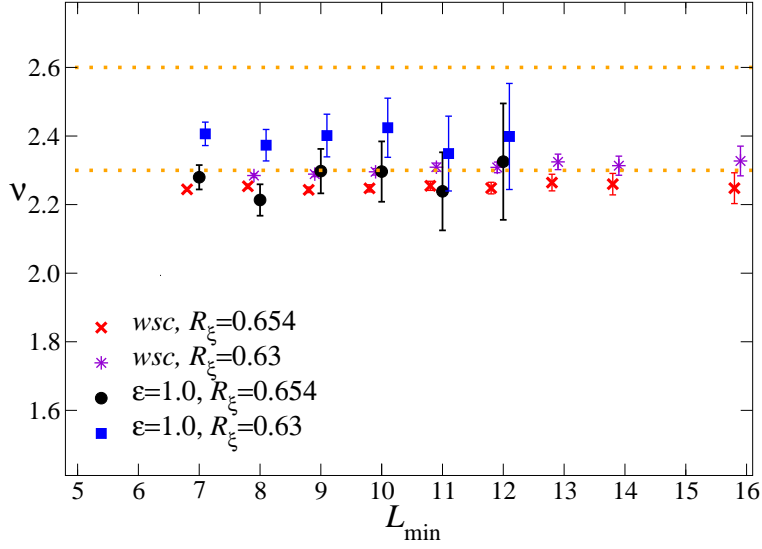}}
\vspace{2mm}
\caption{ (Color online) Estimates of the exponent $\nu$ from the analysis
  of $\bar{U}_4'$, at $R_\xi=0.63$ and $R_\xi=0.654$, 
  for the $\pm J$ Ising model at
  $p=0.5$, obtained by fits without scaling corrections (wsc),
  Eq.~(\ref{ansatznu0}), and fits
  with scaling corrections, Eq.~(\ref{ansatznu}).  
We only show the results of the fits which satisfy 
$\chi^2/{\rm DOF}\lesssim 1.5$,
up to values of $L_{\rm min}$ where the errors blow up.
The dotted lines correspond to $\nu=2.45(15)$.
}
\label{nup0p5u4}
\end{figure}

We estimate the critical exponent $\nu$ from the finite-size behavior of 
$R_\xi'\equiv d R_\xi/d\beta$ and $U_4'\equiv d U_4/d\beta$ at a fixed 
value $R_{\xi,f}$ of $R_\xi$. 
As we discussed in Sec.~\ref{fsssec}, the best choice 
for $R_{\xi,f}$ is $R^*_\xi$, otherwise $\bar{R}_\xi'$ and $\bar{U}_4'$
present analytic corrections. In the present case, 
the estimate (\ref{tcest}) of $R_\xi^*$ is not very precise,
hence the corrections of order $L^{-1/\nu} \sim L^{-0.4}$ cannot be completely
neglected. However, since 
their amplitude is proportional to $(R_{\xi}-R_\xi^*)$, 
we can assume that they are small
for $R_{\xi,f}$ close to $R_\xi^*$. Thus, in order to estimate the systematic 
error induced by them, we proceed as follows. We repeat the analysis 
of $\bar{R}'$ at fixed $R_\xi$ using two different values of $R_\xi$ and 
{\em neglecting} in both cases the $L^{-1/\nu}$ corrections: 
we use $R_\xi = 0.630$ and $R_\xi = 0.654$, which are 
close to our best estimate $R_\xi^* = 0.645(15)$. 
For both values of $R_\xi$ we determine an estimate 
of $\nu$. Our final result is obtained by linear interpolation 
to $R_\xi = 0.645$.  The systematic error is 
estimated from the difference of the results at $R_\xi = 0.630$ and
$R_\xi = 0.645$.

\begin{figure}[tb]
\centerline{\psfig{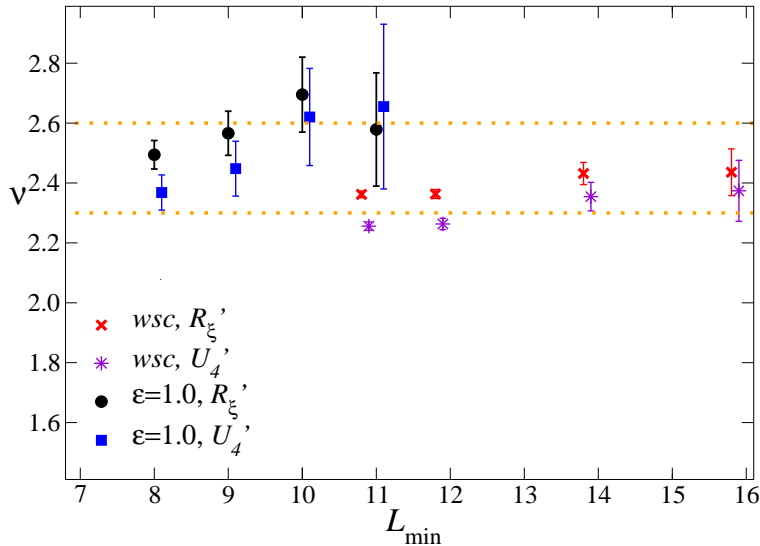}}
\vspace{2mm}
\caption{ (Color online) Estimates of the exponent $\nu$ from the analysis
  of $\bar{R}_\xi'$ and $\bar{U}_4'$, at $R_\xi=0.63$ for the 
  $\pm J$ Ising model at
  $p=0.7$, obtained by fitting the data without scaling corrections (wsc) and 
  with scaling corrections, Eq.~(\ref{ansatznu}).
We only show the results of the fits with $\chi^2/{\rm DOF}\lesssim 1.5$,
up to values of $L_{\rm min}$ where the errors blow up.
The dotted lines correspond to $\nu=2.45(15)$.  }
\label{nup0p763}
\end{figure}

\begin{figure}[tb]
\centerline{\psfig{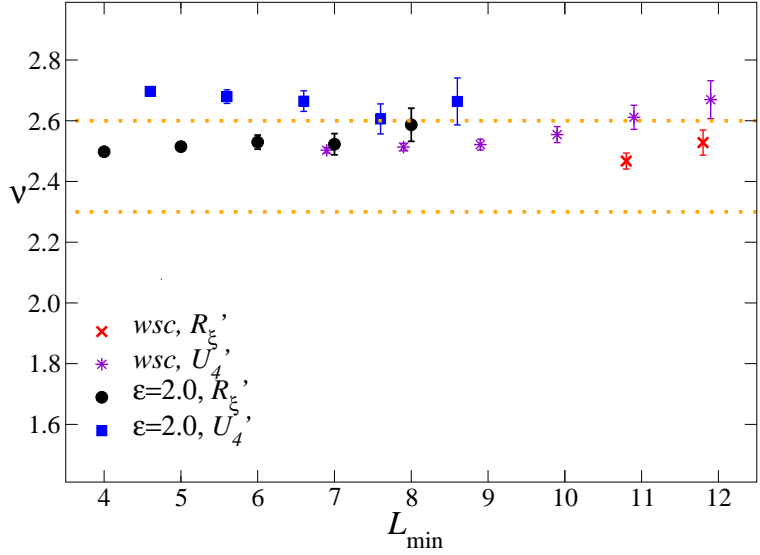}}
\vspace{2mm}
\caption{ (Color online) Estimates of the exponent $\nu$ from the analysis
  of $\bar{R}_\xi'$ and $\bar{U}_4'$, at $R_\xi=0.63$ for the BDBIM
  at $p_b=0.45$, obtained by fitting the data
  without scaling corrections (wsc) and 
  with scaling corrections, Eq.~(\ref{ansatznu}).  
  We only show the results of the fits with reasonable
  values of $\chi^2/{\rm DOF}$, up to values of $L_{\rm min}$ where the errors
  blow up.  The dotted lines correspond to $\nu=2.45(15)$.  }
\label{nudil63}
\end{figure}

The exponent $\nu$ is obtained from fits of 
$\bar{R}_\xi'$ and $\bar{U}_4'$ to
\begin{eqnarray}
&&{\rm ln} \bar{R}' = a + {1\over \nu} {\rm ln} L, \label{ansatznu0}\\
&&{\rm ln} \bar{R}' = a + {1\over \nu} {\rm ln} L + b L^{-\epsilon},
\label{ansatznu}
\end{eqnarray}
with $\epsilon=\omega=1.0(1)$. 
The results of the fits of $\bar{R}_\xi'$ 
for the $\pm J$ Ising model with $p=0.5$ 
are shown in  Fig.~\ref{nup0p5rxi} 
as a function of the minimum lattice size
$L_{\rm min}$ allowed in the fits.  They are quite stable with respect to 
$L_{\rm min}$, depend very little on the fixed value $R_\xi$, 
and change only slightly as $\omega$ 
varies by $\pm 0.1$, corresponding to one error bar.
In particular, the fit of $\bar{R}_\xi'$ at $R_\xi=0.654$
to Eq.~(\ref{ansatznu})
gives $\nu=2.52(4)[2]$
($\nu=2.53(7)[1]$) for $L\ge L_{\rm min}=8$ ($L_{\rm min}=10$); here
the error in brackets takes into
account the uncertainty on $\omega$. Analogously, the fit of the data at
$R_\xi=0.630$ gives 
$\nu=2.51(4)[2]$ ($\nu=2.47(7)[1]$) for the same values of $L_{\rm min}$.
In both cases
$\chi^2/{\rm DOF}\approx 1.1$ ($\chi^2/{\rm DOF}\approx 1.3$). 
Therefore, we obtain $\nu = 2.52(6)$ and $2.51(8)$ at $R_\xi=0.645$
for $L_{\rm min}=8,10$. The systematic error due to the analytic 
corrections is small, 0.01 and 0.04 for the two values of $L_{\rm min}$.
The results from
fits of $\bar{U}_4'$, 
which are shown in Fig.~\ref{nup0p5u4}, favor a smaller value
of $\nu$, although in substantial agreement. Indeed, fits with $L_{\rm min}=10$
give $\nu=2.30(9)$ at $R_\xi=0.654$ and $\nu=2.42(9)$ at $R_\xi=0.630$, thus
suggesting the estimate $\nu=2.35(9)$. The error due to the 
analytic corrections is apparently larger than that obtained in the 
analysis of $\bar{R}'_\xi$, approximately 0.07. As our final result we take
\begin{equation}
\nu=2.45(15),
\label{nuest}
\end{equation}
which is consistent with the results obtained for 
$\bar{U}'_4$ and $\bar{R}'_\xi$.

Results at $R_\xi=0.63$ for the $\pm J$ model at $p=0.7$ and the BDBIM
at $p_b=0.45$, shown in Figs.~\ref{nup0p763} and $\ref{nudil63}$,
respectively, are fully consistent with the estimate (\ref{nuest}). 
For the $\pm J$ model at $p=0.7$,
fits with $\varepsilon=1.0$ and $L_{\rm min}=8$ of $\bar{R}_\xi'$ and 
$\bar{U}_4'$ at fixed $R_\xi = 0.63$ give $\nu=2.49(5)$ and $\nu=2.37(6)$,
respectively. In the case of the BDBIM at $p_b=0.45$ and again for
$R_\xi=0.63$, fits without scaling corrections 
are in full agreement. Fits of $\bar{R}_\xi'$ to (\ref{ansatznu})
assuming $\varepsilon=2.0$ are also consistent: for $L_{\rm min}=7$ they 
give $\nu=2.52(4)$. On the other hand, the fits of $\bar{U}_4'$ 
with $\varepsilon=2.0$
give results that are systematically larger; for instance, 
we obtain $\nu=2.61(5)$ for $L_{\rm min}=7$. This slight discrepancy is 
probably due to the fact that scaling corrections with exponent $\omega = 1$
are small in the BDBIM at $p_b=0.45$ but not completely negligible: 
thus, the fits 
assume that corrections vanish faster than they really do, 
obtaining a slightly incorrect asymptotic estimate.

\subsection{The critical exponent $\eta$}
\label{critexpeta}

\begin{figure}[tb]
\centerline{\psfig{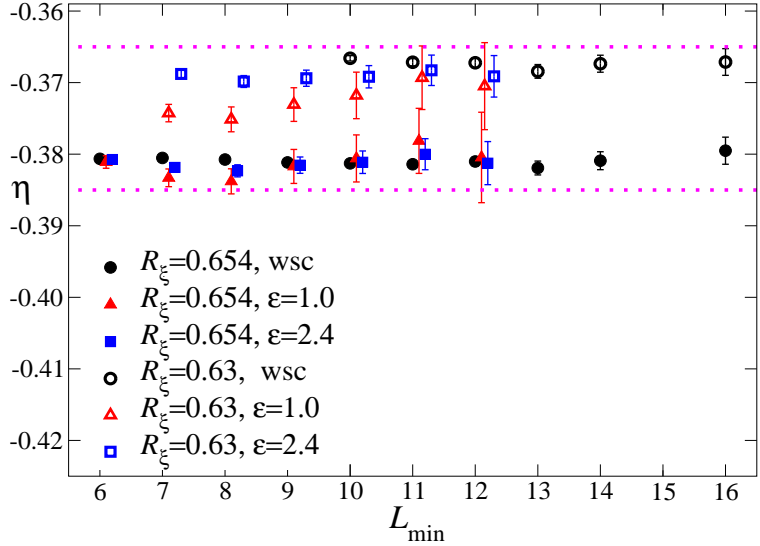}}
\vspace{2mm}
\caption{
  (Color online) Estimates of the exponent $\eta$ from the analysis of
  $\bar{\chi}(L,R_\xi)$
  for $R_\xi=0.63$ and $R_\xi=0.654$, for the $\pm J$ Ising model at
  $p=0.5$.  Only results of fits with $\chi^2/{\rm DOF}<1.5$ are shown.  The
  dotted lines correspond to the estimate $\eta=-0.375(10)$. }
\label{etap0p5}
\end{figure}

\begin{figure}[tb]
\centerline{\psfig{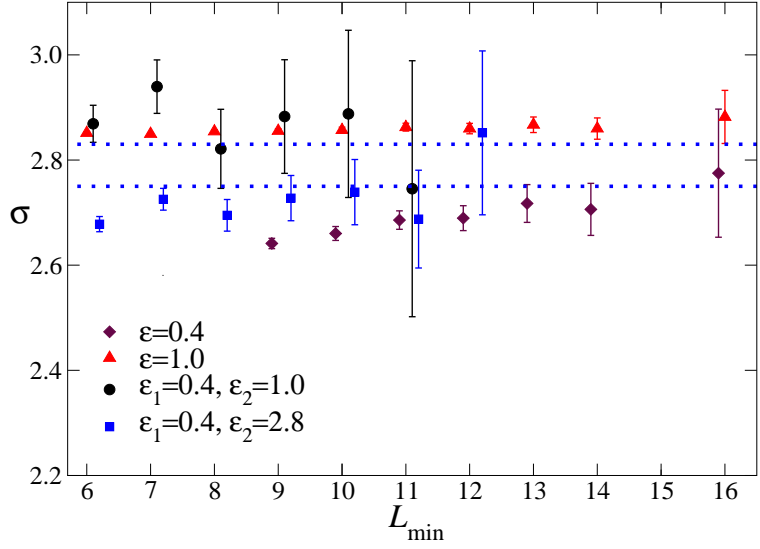}}
\vspace{2mm}
\caption{
  (Color online) Estimates of the critical exponent $\sigma\equiv 1/\nu + 2 -
  \eta$, as obtained by analyzing $\chi'$ at $R_\xi=0.654$.  The
  dotted lines correspond to $\sigma=2.78(4)$.  }
\label{chipp0p5}
\end{figure}

We estimate the exponent $\eta$ by analyzing the susceptibility $\chi$ at
fixed $R_\xi$.  Its critical behavior is discussed in Sec.~\ref{fsssec}:
at fixed $R_\xi$, it behaves as  
\begin{equation}
\bar{\chi}(L,R_\xi) = 
  a L^{2-\eta} \left[ 1 + a_a (R_\xi-R_\xi^*) L^{-1/\nu} + \cdots +
a_\omega L^{-\omega} + \ldots + a_b L^{-2+\eta} + \cdots \right],
\label{chicr}
\end{equation}
where the $O(L^{-2+\eta})$ term is the background
contribution, cf. Eq.~(\ref{chiexp_1}).  Since $R_\xi$ is not exactly
equal to $R_\xi^*$, we must take into account 
the $O(L^{-1/\nu})$ corrections. As for $\nu$,
the systematic error they induce is estimated by comparing 
the results at $R_\xi = 0.630$ and $R_\xi = 0.654$, which are 
close to the best estimate $R_\xi^*=0.645(15)$.  We
perform fits with and without scaling corrections, i.e. to
\begin{eqnarray}
&&{\rm ln} \chi = c + (2-\eta) {\rm ln} L , \label{chiwsc} \\
&&{\rm ln} \chi = c + (2-\eta) {\rm ln} L + c_\varepsilon L^{-\varepsilon},
\label{chico}
\end{eqnarray}
with $\varepsilon=1$ (the leading scaling correction arising from irrelevant
scaling fields) and $\varepsilon=2.4\approx 2-\eta$ (to check the 
relevance of the background term, which might be large for small lattices)
at $R_\xi=0.63$ and $R_\xi=0.654$.  Fig.~\ref{etap0p5} shows 
the estimates of $\eta$ for the $\pm J$ model at $p=0.5$. 
As already mentioned, the systematic error due to the
neglected $L^{-1/\nu}$ corrections is estimated from the
difference of the estimates at $R_\xi=0.63$ and $R_\xi=0.654$.  Following the
same method used to determine $\nu$, we obtain $\eta = -0.375(2)\{8\}$, where
the first error is the statistical one while the second error takes into
account the systematic error due to the residual $O(L^{-1/\nu})$ corrections
and corresponds to the uncertainty on our estimate of $R_\xi^*$.  Slightly
smaller but perfectly consistent results are obtained in the analyses of the
data for the other models.  For example, in the case of the $\pm J$ Ising
model at $p=0.7$, a fit of the data at $R_\xi=0.63$
to Eq.~(\ref{chico}) with $\varepsilon=1.0$ gives
$\eta=-0.381(5)$ for $L\ge L_{\rm min}=10$ 
($\chi^2/{\rm DOF}\approx 1.6$).  In the case of the
BDBIM at $p_b=0.45$, a fit of the data at $R_\xi=0.63$ to
Eq.~(\ref{chico}) with $\varepsilon=2.0$ gives
$\eta=-0.385(4)$ for $L\ge L_{\rm min}=9$ ($\chi^2/{\rm DOF}\approx 1.4$). 
Taking also these
results into account, we arrive at the final estimate
\begin{equation}
\eta = -0.375(10).
\label{etaest}
\end{equation}
We finally discuss the behavior of the derivative $\chi'\equiv d\chi/ d \beta$
of the susceptibility, which in some numerical works, see, e.g.
Ref.~\onlinecite{KKY-06}, has led to contradictory results for the exponent
$\nu$.  We show in the following that this discrepancy is essentially due to
the analytic $O(L^{-1/\nu})$ corrections discussed 
in Sec.~\ref{fsssec}, cf. Eq.~(\ref{chipbc}).  At fixed
$R_\xi$, $\bar{\chi}'$ is expected to behave as
\begin{equation}
\bar{\chi}'(L,R_\xi) = b L^\sigma \left[ 1 + b_a L^{-1/\nu} + \cdots +
b_\omega L^{-\omega} + \ldots + b_b L^{-\sigma} + \cdots \right],
\label{chipcr}
\end{equation}
where $\sigma\equiv {1/\nu+2-\eta}$. Using our best estimates of $\nu$
and $\eta$, we obtain $\sigma=2.78(4)$.  Unlike the case of $\chi$ and the
derivative of the phenomenological quantities, the amplitude $b_a$ of the
$O(L^{-1/\nu})$ corrections does not vanish at $T_c$ and thus at
$R_\xi=R_\xi^*$; therefore, it is not expected to be small
when $R_\xi\approx R_\xi^*$.  The $O(L^{-\sigma})$ term in 
Eq.~(\ref{chipcr})
is the background contribution, cf. Eq.~(\ref{chidexp_1}).  In
Fig.~\ref{chipp0p5} we show the estimates of $\sigma$ obtained in 
fits of $\ln \chi'$ to 
\begin{eqnarray}
&&{\rm ln} \chi' = c + \sigma {\rm ln} L + c_\varepsilon L^{-\varepsilon}, \nonumber \\
&&{\rm ln} \chi' = c + \sigma {\rm ln} L + c_1 L^{-\varepsilon_1}
+ c_2 L^{-\varepsilon_2},
\label{chipfits}
\end{eqnarray}
for several values of $\varepsilon,\varepsilon_1,\varepsilon_2$.
The results are consistent with the expected value $\sigma\approx 2.78$,
when the $O(L^{-{1/\nu}})$ corrections are taken into account.

\section{High-temperature estimates} 
\label{HT-est}

In the parallel-tempering simulations we have collected a large 
amount of data for several values of $\beta$ in the high-temperature
phase. They have not been used in the analyses we have presented in the 
previous Sections. Here, we shall present analyses that consider 
all high-temperature results.
They fully confirm the critical-point estimates discussed above
and provide additional evidence that all models belong to the same 
universality class.

\subsection{Finite-size scaling behavior of $\xi$ and estimates of 
$\nu$}

\begin{figure}[tb]
\centerline{\psfig{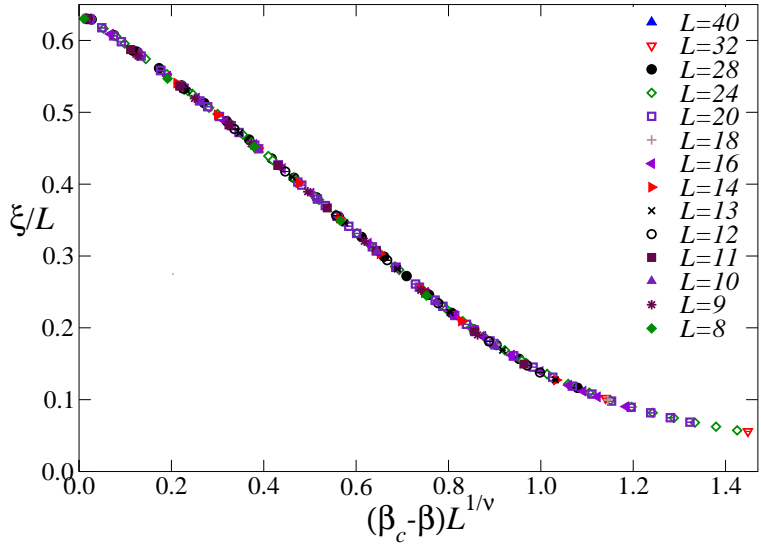}}
\vspace{4mm}
\caption{
  (Color online) Plot of $\xi/L$ vs $(\beta_c - \beta) L^{1/\nu}$ with
  $\beta_c = 0.902$ and $\nu = 2.45$. Data for the 
  $\pm J$ model with $p = 0.5$.
  }
\label{xil-HT}
\end{figure}

We determine $\nu$ from the FSS behavior of the correlation length.
The starting point is the FSS equation 
\begin{equation}
{\xi(\beta,L)\over L} = f(u_t L^{1/\nu}) + 
    {v_\omega(\beta) \over L^{\omega}} f_\omega(u_t L^{1/\nu}) + \cdots , 
\label{FSS-xi-HT}
\end{equation}
where $f(x)$ and $f_\omega(x)$ are universal 
(once the normalization of the scaling fields has been fixed) and 
have a regular expansion in powers of $x$.
In order to ensure a finite infinite-volume limit, for $x\to\infty$
they behave as 
\begin{equation}
f(x) \sim x^{-\nu}, \qquad\qquad f_\omega(x) \sim x^{\nu(\omega-1)}.
\label{large-xf}
\end{equation}
The scaling field $u_t$ is an analytic function of $\beta$ that vanishes
at the critical point. Hence, it has an expansion 
\begin{equation}
u_t = \beta_c - \beta + b (\beta_c - \beta)^2 + 
    O[(\beta_c - \beta)^3].
\label{expan-ut}
\end{equation}
In Fig.~\ref{xil-HT} we report $\xi(\beta,L)/L$ versus $(\beta_c - \beta)
L^{1/\nu}$ using the data for the $\pm J$ model at $p=0.5$ and the estimates
of $\beta_c$ and $\nu$ obtained in the previous Sections: $\beta_c = 0.902$
and $\nu=2.45$. The data collapse quite precisely onto a single curve. This
indicates that the nonanalytic scaling corrections are small and so are the
analytic ones due to the nontrivial dependence of $u_t$ on $\beta$.

In order to obtain quantitative estimates of $\nu$, we follow
Ref.~\onlinecite{CMPV-03} and perform three different fits of ${\xi(\beta,L)/
  L}$.  In the first fit we neglect the nonanalytic scaling corrections, set
$u_t = \beta_c - \beta$, and fit the data to (fit a)
\begin{equation}
{\xi(\beta,L)\over L} = 
   P_n(x)^{-\nu/n}, 
  \qquad \qquad x \equiv (\beta_c - \beta) L^{1/\nu},
\label{fita}
\end{equation}
where $P_n(x)$ is a polynomial of degree $n$. We have chosen polynomials
for computational convenience, but any other complete set of functions 
can be used.
Note that the specific choice (\ref{fita})
of fitting function satisfies the large-$x$ behavior (\ref{large-xf}). In this
fit, the free parameters are the $(n+1)$ coefficients of the polynomial
$P_n(x)$, the critical inverse temperature $\beta_c$, and
$\nu$.\cite{footnote-fita} The critical-point value $R_\xi^*$ is equal to
$P_n(x=0)^{-\nu/n}$. 

In order to take into account the analytic corrections (fit b), we slightly
modify the Ansatz (\ref{fita}), setting $x \equiv \beta_c - \beta + b (\beta_c
- \beta)^2$ and taking $b$ as an additional free parameter.  Finally, we
consider the nonanalytic scaling corrections.  We use the Ansatz (fit c)
\begin{equation}
{\xi(\beta,L)\over L} =
  \left[ P_n(x) + {1\over L^\omega} (1 + a x)^{\omega\nu} Q_n(x)
  \right]^{-\nu/n},
  \qquad \qquad x \equiv (\beta_c - \beta) L^{1/\nu},
\label{fitc}
\end{equation}
where $P_n(x)$ and $Q_n(x)$ are both polynomials of degree $n$, and $a$ is a
free parameter. We can check that Ansatz (\ref{fitc}) has the correct
infinite-volume limit.  For $L\to\infty$ at fixed $\beta$, i.e., for
$x\to\infty$, we have $P_n(x) \approx p_n x^n$, $Q_n(x) \approx q_n x^n$, and
\begin{eqnarray}
\xi(\beta,L) &\approx& L
   \left[ p_n x^n + {1\over L^\omega} (a x)^{\omega\nu} q_n x^n
  \right]^{-\nu/n} 
   \approx L x^{-\nu} \left[ p_n + 
        a^{\omega\nu} q_n (\beta_c - \beta)^{\omega\nu} \right]^{-\nu/n}
\nonumber \\
   &\sim&  (\beta_c - \beta)^{-\nu} 
     \left[1 + k (\beta_c - \beta)^{\Delta}\right] 
\end{eqnarray}
where $\Delta \equiv  \omega\nu$.
In these fits $n$ must be chosen so that $P_n$ provides 
an accurate parametrization of the scaling function. 
We have tried several values of
$n$, until the $\chi^2$ of the fit does not change significantly as $n$
increases by 1. In practice, we have taken $n$ between 6 and 10. In fit
(\ref{fitc}) corrections to scaling are parametrized by a polynomial that has
the same degree as that parametrizing the leading behavior. Our data are not
so precise and corrections are not so large to require such a large number of
parameters. To reduce the number of free parameters we have taken $n=6,9$ and
set
\begin{equation}
Q_n(x) = q_0 + q_3 x^3 + \cdots + q_n x^n.
\end{equation}
This choice, which is completely arbitrary, significantly reduces the number
of free parameters, but still allows us to parametrize accurately (at the
level of the statistical errors) the scaling corrections.

\begin{figure}[tb]
\centerline{\psfig{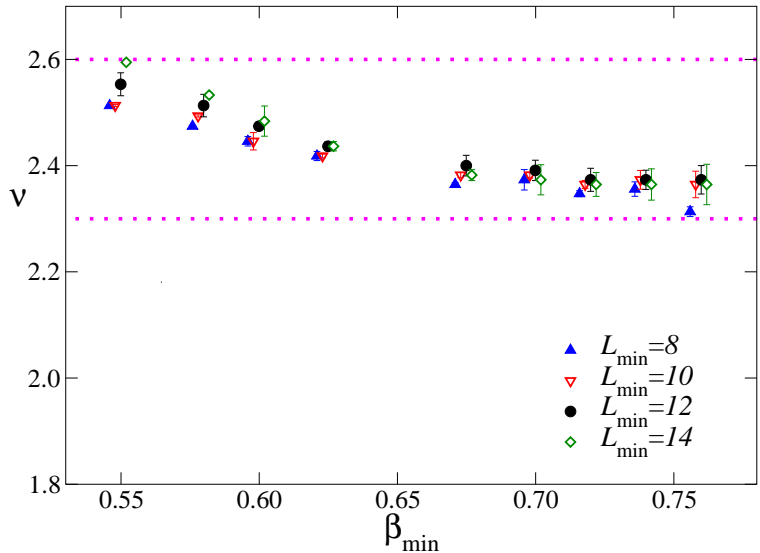}}
\vspace{4mm}
\caption{
  (Color online) Estimates of $\nu$ for the
  $\pm J$ model at $p = 0.5$.
  We report estimates obtained by fitting $\xi/L$ to 
  the Ansatz (\ref{fita})  for several values 
  of $L_{\rm min}$ and $\beta_{\rm min}$.
  The dotted lines correspond to the estimate $\nu = 2.45(15)$.
  }
\label{nu-HT1}
\end{figure}

In order to take into account the additional scaling corrections which are not
taken into account by our fit Ans\"atze, we have repeated each fit several
times; each time we only include data such that $\beta\ge \beta_{\rm min}$ and
$L\ge L_{\rm min}$.
In Fig.~\ref{nu-HT1} we report the estimates of $\nu$ obtained in fits of
$\xi/L$ to Eq.~(\ref{fita}) for the $\pm J$ model at $p=0.5$. Corrections to
scaling are clearly visible for small values of $\beta_{\rm min}$ and $L_{\rm
  min}$.  Indeed, at fixed $L_{\rm min}$ the estimates decrease, becoming
approximately independent of $\beta_{\rm min}$ for $\beta_{\rm min} \gtrsim
0.65$. The dependence on $L_{\rm min}$ is instead tiny, and all results with
$L_{\rm min} \ge 10$ are consistent within errors.

\begin{figure}[tb]
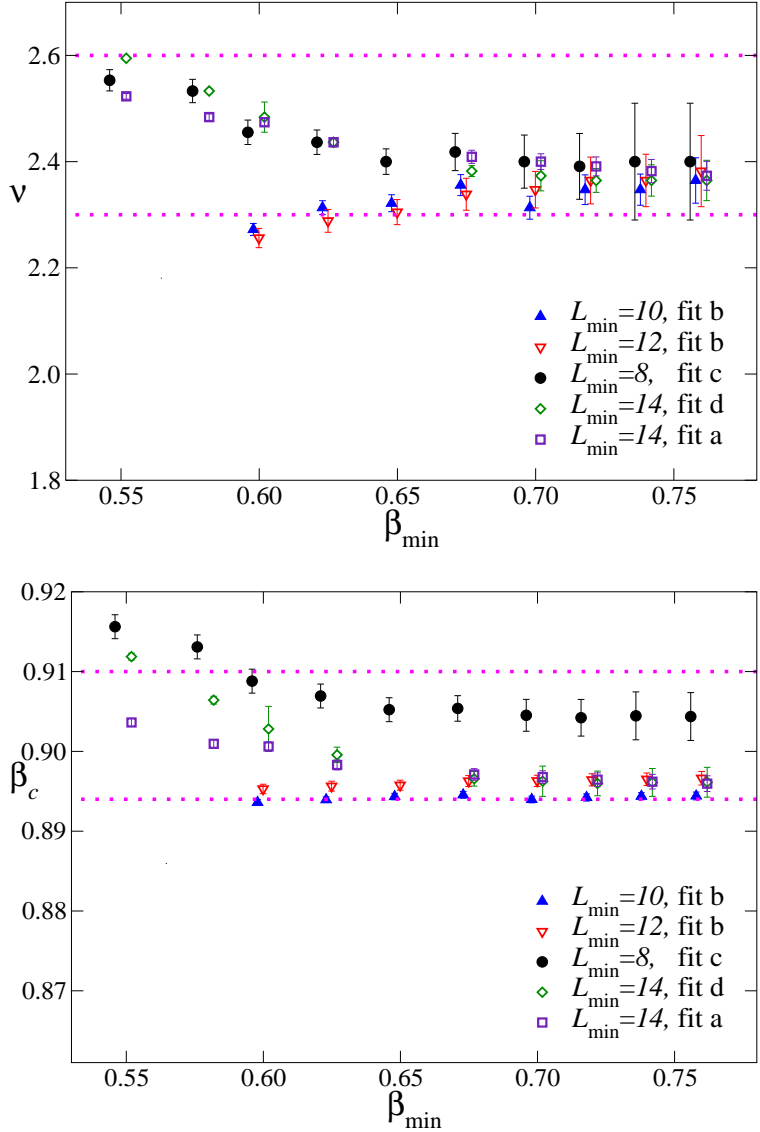

\centerline{\psfig{width=10truecm,angle=0,file=fig19a.eps}}
\vspace{4mm}
\centerline{\psfig{width=10truecm,angle=0,file=fig19b.eps}}
\vspace{2mm}
\caption{
  (Color online) Estimates of $\nu$ (top) and $\beta_c$ (bottom) for the 
  $\pm J$ model at $p = 0.5$. 
  We report estimates obtained by fitting $\xi/L$ to 
  Eq.~(\ref{fita}) (fit a), to Eq.~(\ref{fita}) 
  with $x \equiv [(\beta_c - \beta) + b (\beta_c - \beta)^2] L^{1/\nu}$ 
  (fit b), to Eq.~(\ref{fitc}) with $\omega = 1$ (fit c),
  and to the extended-scaling 
  Ansatz (\protect\ref{fitCPB}) (fit d). 
  The dotted lines correspond to the estimates $\nu = 2.45(15)$ (top) and 
  $\beta_c = 0.902(8)$ (bottom).
  }
\label{nu-betac-HT}
\end{figure}

The results of fits that also take into account the analytic and the
nonanalytic corrections (with $\omega = 1$) are reported in
Fig.~\ref{nu-betac-HT}.  As far as $\nu$ is concerned, no significant
differences are observed and in all cases the results become independent of
$\beta_{\rm min}$ for $\beta_{\rm min} \gtrsim 0.65$. 
All results (with their statistical
errors) lie in the interval $2.3\le \nu \le 2.5$, and are therefore in perfect
agreement with the estimate $\nu = 2.45(15)$ obtained before.  Corrections to
scaling are more evident in the analyses of $\beta_c$. Indeed, while analyses
without nonanalytic scaling corrections give estimates of $\beta_c$ that
cluster around 0.895, those that take the corrections into account give values
which are somewhat larger. In any case, all results are consistent with the
estimate $\beta_c = 0.902(8)$ obtained in Sec.~\ref{crittemp1}.
Finally, these analyses also
provide estimates of $R^*_\xi$.  Analyses without nonanalytic scaling
corrections give $R^*_\xi = 0.632(5)$, while those which include scaling
corrections give a somewhat higher value $R^*_\xi = 0.648(5)$. Also in this
case, scaling corrections appear to be quite relevant. These results are
perfectly consistent with the estimate (\ref{tcest}), $R^*_\xi = 0.645(15)$.
The analyses that take into account the analytic scaling corrections (fit b)
also give estimates of the constant $b$ that appears in the expansion
(\ref{expan-ut}) of $u_t$: they vary somewhat with $\beta_{\rm min}$ and give
approximately $0\lesssim b \lesssim 0.3$.

In order to verify universality, we compute $\nu$ 
and determine the FSS curves also for the $\pm J$ model at $p=0.7$ 
and the BDBIM at $p_b = 0.45$. We report here only estimates of $\nu$ obtained 
by fitting the data to Eq.~(\ref{fita}), since we do not have 
data precise enough to allow for a detailed study of the scaling 
corrections. In any case, the results for $p = 0.5$ indicate that 
scaling corrections do not play much role in the determinations of $\nu$.
For the $\pm J$ model at $p = 0.7$ we obtain: 
$\nu = 2.382(5)$ [$\nu = 2.427(10)$] for $\beta_{\rm min} = 0.59$ and 
$L_{\rm min} = 7$ [$L_{\rm min} = 9$]; 
$\nu = 2.347(10)$ [$\nu = 2.382(20)$] for $\beta_{\rm min} = 0.68$ and the 
same values of $L_{\rm min}$.
For the BDBIM at $p_b = 0.45$ we obtain:
$\nu = 2.574(6)$ [$\nu = 2.637(7)$] for $\beta_{\rm min} = 0.82$ and again
$L_{\rm min} = 7$ [$L_{\rm min} = 9$]; 
$\nu = 2.409(12)$ [$\nu = 2.437(18)$] for $\beta_{\rm min} = 1.02$.
These results are in very good agreement with the estimate 
$\nu = 2.45(15)$ reported above.

\begin{figure}[tb]
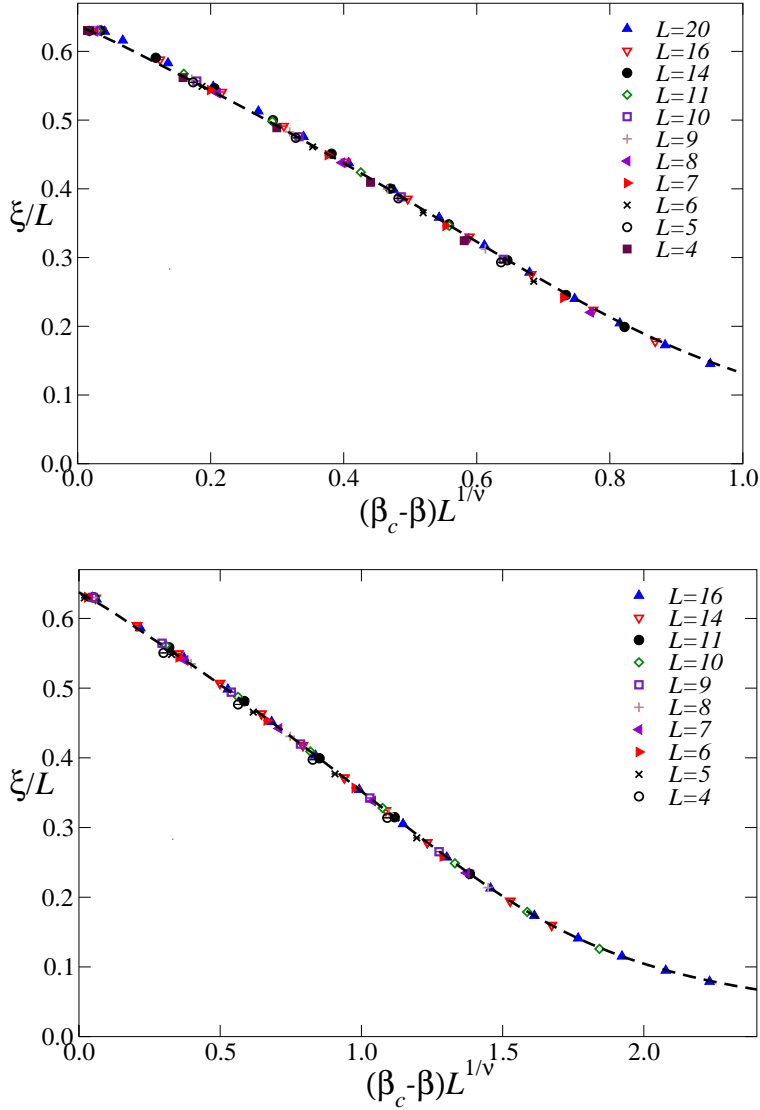

\centerline{\psfig{width=10truecm,angle=0,file=fig20a.eps}} 
\vspace{4mm}
\centerline{\psfig{width=10truecm,angle=0,file=fig20b.eps}}
\vspace{2mm}
\caption{
  (Color online) Plot of $\xi/L$ vs $(\beta_c - \beta) L^{1/\nu}$
   for the $\pm J$ model at $p=0.7$ (top) and for the BDBIM at $p_b = 0.45$
   (bottom). We use $\nu = 2.45$, $\beta_c = 0.87$ (top), and 
   $\beta_c = 1.54$ (bottom).
   The dashed line 
   is the curve $R_\xi(cx)$, where $R_\xi(x)$ is the function
   (\ref{fxiFSS}), which is estimated in fits of $\xi/L$ for  
   the $\pm J$ model at $p = 0.5$, and $c$ is a model-dependent constant:
   $c \approx 1.026$ for the $\pm J$ model at $p=0.7$ and 
   $c \approx 0.5641$ for the BDBIM at $p_b = 0.45$.
  }
\label{FSSxi-altrimodelli}
\end{figure}

In order to verify the universality of the FSS curves we first fitted the data
for the $\pm J$ model at $p = 0.5$ presented in Fig.~\ref{xil-HT}. Taking $\nu
= 2.45$, $\beta_c = 0.902$, and using only the data satisfying $L\ge 10$,
$\beta \ge 0.62$, we obtain
\begin{equation}
   {\xi\over L} = R_\xi(x) \qquad\qquad x \equiv (\beta_c - \beta) L^{1/\nu},
\end{equation}
with (this expression is valid in the interval of values of $x$ 
for which we have data, i.e., for $0\le x \lesssim 1.5$)
\begin{eqnarray}
R_\xi(x) &=& \left(
   6.2828 + 16.8612 x -
          39.3317 x^2 + 1926.5102 x^3 \right.
\nonumber \\
       &&  - 17659.3388 x^4 + 88711.1141 x^5 -
          256918.2481 x^6 + 446776.5137 x^7 
\nonumber  \\
       && \left. - 452723.8074 x^8 + 243001.4960 x^9 -
          50040.5243 x^{10}\right)^{-0.245}.
\label{fxiFSS}
\end{eqnarray}
The function $R_\xi(x)$ is universal apart from a rescaling of its argument.
Thus, if we plot $\xi/L$ vs $x\equiv (\beta_c - \beta) L^{1/\nu}$
in any model, the data should fall on the curve $R_\xi(c x)$, where $c$
is a model-dependent constant. In Fig.~\ref{FSSxi-altrimodelli}
we report the data for the $\pm J$ model at $p = 0.7$ and the BDBIM
at $p_b = 0.45$. The results show very good scaling and fall on top
of the curve computed from the data of the bimodal model at $p = 0.5$, 
confirming universality.

\begin{figure}[tb]
\centerline{\psfig{width=10truecm,angle=-90,file=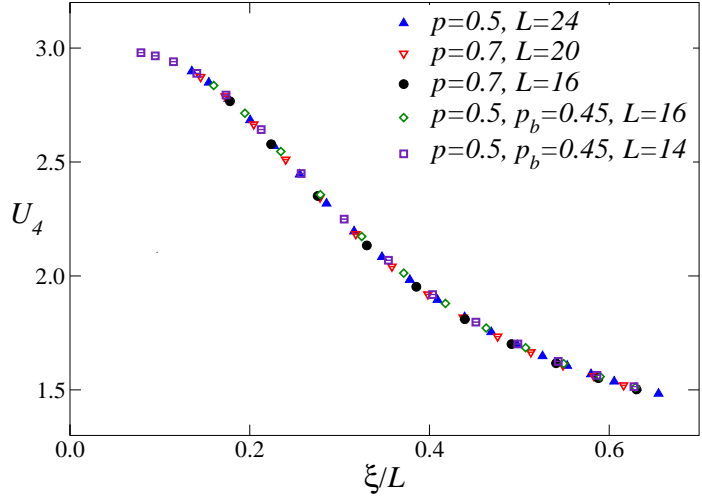}} 
\caption{
  (Color online) Plot of $U_4$ vs $\xi/L$ for the $\pm J$ model at $p =0.5$
  and 0.7 and the BDBIM at $p_b = 0.45$.
  }
\label{fig:Ux}
\end{figure}

We finally consider the cumulant $U_4$. In the 
critical limit $U_4$ should be a universal function of $\xi/L$, independent
of the model. Corrections scale as $L^{-\omega} h(\xi/L)$, 
where $h(x)$ is also universal, apart from a multiplicative constant.
The numerical estimates of $U_4$ and $\xi/L$ are reported 
in Fig.~\ref{fig:Ux}. We consider the $\pm J$ model at $p=0.5$ and $p=0.7$
and the BDBIM at $p_d = 0.45$. We only consider the largest lattices,
so that nonanalytic scaling corrections are not visible on the scale
of the figure (a detailed study of the $L^{-\omega}$ corrections
for $\xi/L = 0.63$ is reported in Sec.~\ref{unisevi}). All points fall 
quite precisely onto a single curve, confirming that the PG transition in
these three models belongs to the same universality class.

\subsection{Finite-size scaling of the susceptibility and 
estimates of $\eta$}

\begin{figure}[tb]
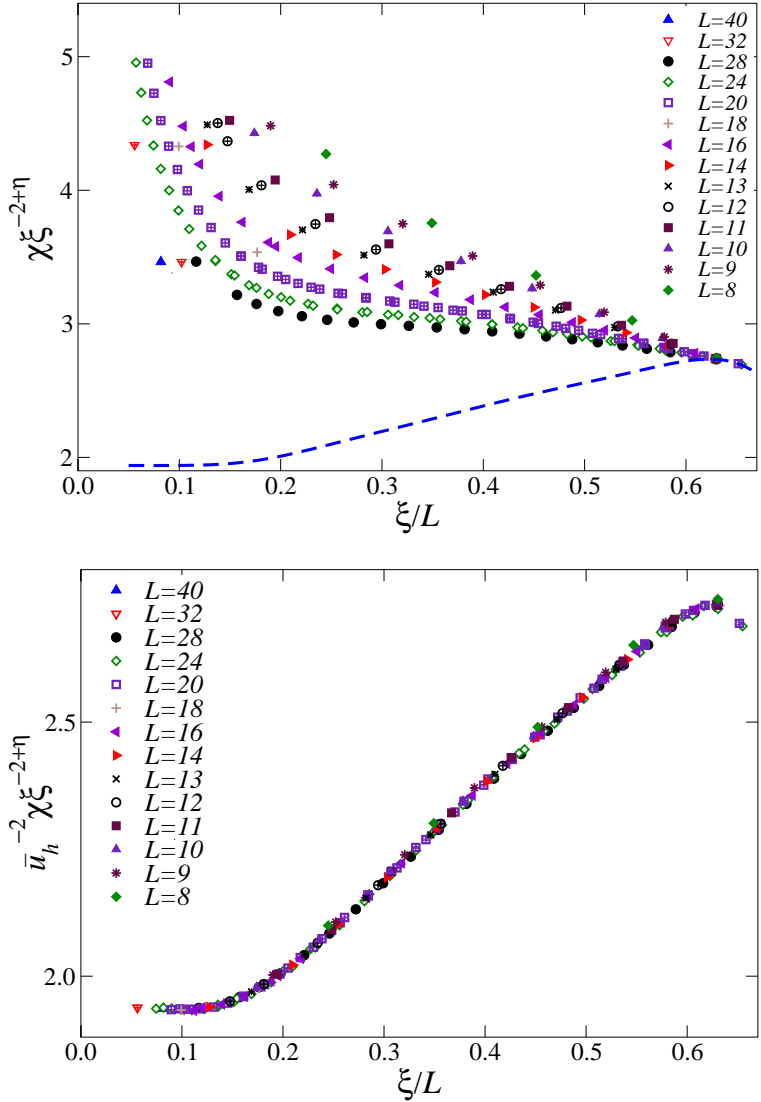

\centerline{\psfig{width=10truecm,angle=0,file=fig22a.eps}} 
\vspace{4mm}
\centerline{\psfig{width=10truecm,angle=0,file=fig22b.eps}}
\vspace{2mm}
\caption{
  (Color online) Plot of $\chi \xi^{\eta - 2}$ (upper panel) 
   and of $\chi \xi^{\eta - 2} \tilde{u}_h^{-2}$ (lower panel) 
   vs $\xi/L$ with $\eta = -0.375$. 
   The dashed line in the upper panel 
   is the universal curve $\tilde{C}(\xi/L)$, as 
   estimated in fits of $\chi$ to Eq.~(\ref{fita-chi}).
   Data for the $\pm J$ model at $p = 0.5$.
  }
\label{chi-HT}
\end{figure}

We now turn to the determination of the critical exponent $\eta$. The starting 
point is Eq.~(\ref{chiexp_1}), which we rewrite as
\begin{equation}
\chi(\beta,L) = L^{2-\eta} \bar{u}_h(\beta)^2 g(u_t L^{1/\nu}) 
   \left[1 + {v_\omega(\beta)\over L^\omega} 
              g_\omega(u_t L^{1/\nu}) + \ldots \right].
\label{chiFSS-HT}
\end{equation}
This equation is not very convenient since it involves $u_t$, hence the 
critical temperature, and the exponent $\nu$. To reduce the number of 
unknown parameters, we note that Eq.~(\ref{FSS-xi-HT}) can be inverted to give
\begin{equation}
u_t L^{1/\nu} = F(\xi/L) + {v_\omega(\beta)\over L^\omega} 
                F_\omega(\xi/L) + \ldots
\end{equation}
Inserting in Eq.~(\ref{chiFSS-HT}) we obtain the scaling form 
\begin{equation}
\chi(\beta,L) = \xi^{2-\eta} \bar{u}_h(\beta)^2 
   C(\xi/L)  \left[1 + v_\omega(\beta) \xi^{-\omega}
              C_\omega(\xi/L) + \ldots \right].
\label{chiFSS-HT1}
\end{equation}
In Eq.~(\ref{chiFSS-HT1}) we have singled out $\xi^{2-\eta}$ and 
$\xi^{-\omega}$ instead of $L^{2-\eta}$ and $L^{-\omega}$. 
With this choice $C(x)$ and $C_\omega(x)$ are regular for $x\to 0$. 
Note the presence of the function $\bar{u}_h(\beta)$. In the FSS limit
$\xi\to \infty$, $L\to \infty$, at fixed $\xi/L$, we always have 
$\beta \to \beta_c$, so that asymptotically it should be possible 
to replace $\bar{u}_h(\beta)$ with the constant $\bar{u}_h(\beta_c)$. 
Therefore, this function gives rise to scaling corrections, that we have 
named analytic corrections in the previous sections. In order to understand 
their relevance for our data, in Fig.~\ref{chi-HT} (upper panel) we plot 
$\xi^{\eta-2} \chi$ versus $\xi/L$ for the 
$\pm J$ model at $p=0.5$. It is evident that the data do not 
fall onto a single curve: the scaling-field term $\bar{u}_h(\beta)$
varies significantly with $\beta$ and therefore cannot be neglected.

The previous discussion indicates that, in order to estimate accurately
the exponent $\eta$, it is essential to include the analytic 
corrections in the fitting function. We perform two fits. In the first one
(fit a) we neglect the nonanalytic scaling corrections and consider
\begin{equation}
\ln {\chi\over \xi^2} = - \eta \ln \xi + P_n(\xi/L) + Q_m(\beta),
\label{fita-chi}
\end{equation}
where $P_n(x)$ and $Q_m(x)$ are polynomials of degree $n$ and $m$,
respectively. Moreover, we require $Q_m(0) = 0$ in order to avoid 
the presence of two constant terms. As before, $n$ and $m$ are varied 
till the quality of the fit does not change significantly by varying the 
parameters by 1. In practice, we take $6\le n,m\le 10$. To include the
scaling corrections, we also consider 
\begin{equation}
\ln {\chi\over \xi^2} = - \eta \ln \xi + P_n(\xi/L) + Q_m(\beta) + 
   \xi^{-\omega} S_p(\xi/L),
\label{fitb-chi}
\end{equation}
where $S_p(x)$ is a polynomial of degree $p$: we take $p\le 3$.
Note that here, as we already did in the analysis of $\xi/L$, we 
neglect the $\beta$ dependence of the scaling field $v_\omega$. 

\begin{figure}[tb]
\centerline{\psfig{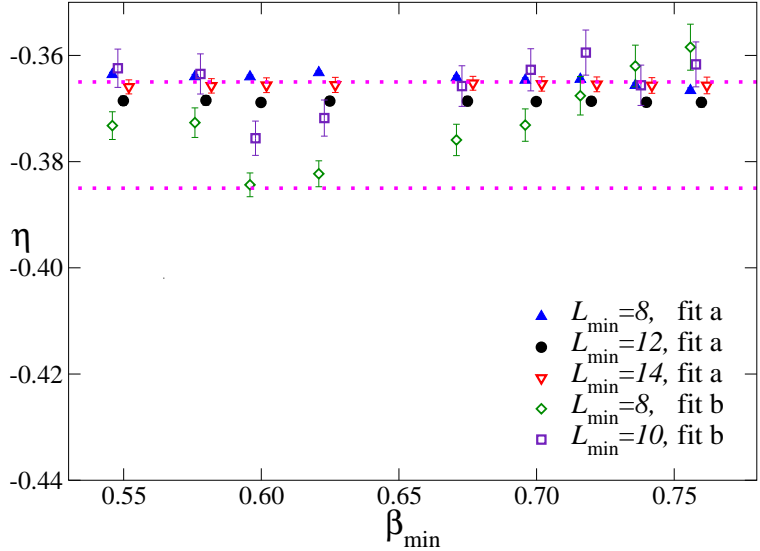}}
\vspace{4mm}
\caption{
  (Color online) Estimates of $\eta$ for the
  $\pm J$ model at $p = 0.5$.
  We report estimates obtained by fitting $\chi$ to
  the Ansatz (\ref{fita-chi}) (fit a) and
  (\ref{fitb-chi}) (fit b)  for several values
  of $L_{\rm min}$ and $\beta_{\rm min}$.
  The dotted lines correspond to the estimate $\eta = -0.375(10)$.
  }
\label{eta-HT}
\end{figure}

\begin{figure}[tb]
\centerline{\psfig{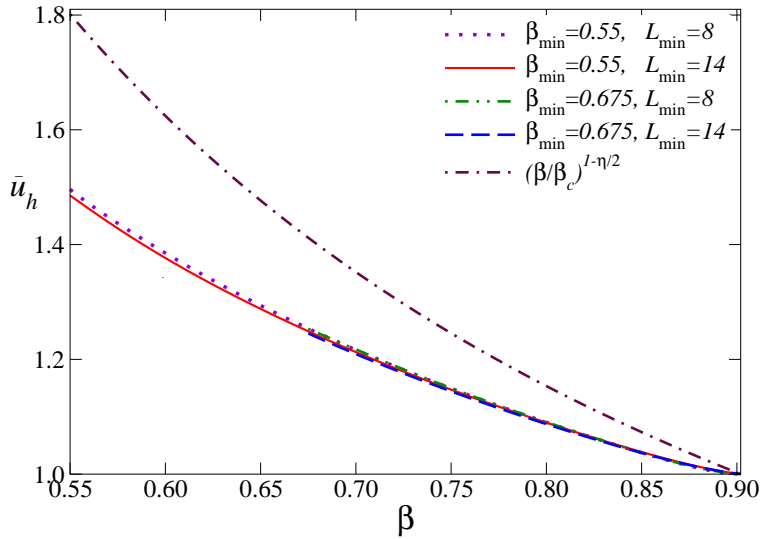}}
\vspace{4mm}
\caption{
  (Color online) Scaling-field function $\tilde{u}_h(\beta)$ 
  for the $\pm J$ model at $p = 0.5$, 
  as determined in the fits of $\chi$ to
  the Ansatz (\ref{fita-chi}), for different values of 
  $\beta_{\rm min}$ and $L_{\rm min}$. 
  It is normalized such that $\tilde{u}_h(\beta_c)=1$. 
  We also report the approximation 
  $\tilde{u}_h(\beta) \approx (\beta/\beta_c)^{1-\eta/2}$,
  which is used in the extended-scaling scheme
  (see the discussion in Sec.~\ref{secVIII.C}).
  }
\label{uh-fun}
\end{figure}

The results of the fits for the 
$\pm J$ model at $p=0.5$ are reported in Fig.~\ref{eta-HT}. The estimates 
of the fit to Eq.~(\ref{fita-chi}) are very stable and show a very
tiny dependence on $\beta_{\rm min}$ and $L_{\rm min}$.
For instance, for $\beta_{\rm min} = 0.55$ we obtain
$\eta = -0.3636(8)$ ($L_{\rm min} = 8$),
$\eta = -0.3659(27)$ ($L_{\rm min} = 14$),
while for $\beta_{\rm min} = 0.75$ we have
$\eta = -0.3666(10)$ ($L_{\rm min} = 8$),
$\eta = -0.3657(31)$ ($L_{\rm min} = 14$). 
Fits with nonanalytic scaling corrections are less stable. 
We observe significant fluctuations that indicate that the data 
are not precise enough to be sensitive to this type of scaling corrections.
This is consistent with what is observed at the critical point: 
While the results for $\eta$ depend strongly on the chosen value for 
$R_{\xi,f}$, indicating that the analytic scaling corrections are important, 
essentially no dependence is observed on the nonanalytic ones;
see, e.g., Fig.~\ref{etap0p5}. In any case, all results are 
consistent with the estimate $\eta = -0.375(10)$ obtained 
in Sec.~\ref{critexpeta}.

The fits also give estimates of the function $\bar{u}_h(\beta)$
that appears in Eqs.~(\ref{chiFSS-HT}) and (\ref{chiFSS-HT1}). In 
Fig.~\ref{uh-fun} we plot 
$\tilde{u}_h(\beta) = \bar{u}_h(\beta)/\bar{u}_h(\beta_c)$ [$\tilde{u}_h(\beta)$
is normalized so that $\tilde{u}_h(\beta_c)=1$] as obtained in the different 
fits. The results corresponding to different values of 
$L_{\rm min}$ and $\beta_{\rm min}$ agree nicely, supporting the 
scaling Ansatz (\ref{chiFSS-HT1}). A simple expression which reproduces the 
results reported in Fig.~\ref{uh-fun} is 
\begin{equation}
\tilde{u}_h(\beta) = 
    1 + 0.556247 (1 - \beta/0.902) + 1.83322 (1 - \beta/0.902)^2,
\end{equation}
which is valid for $0.55\le \beta \le 0.902$.  Once $\tilde{u}_h(\beta)$ has
been determined, we can compute the scaling function $\tilde{C}(\xi/L) =
\bar{u}_h(\beta_c)^2 {C}(\xi/L)$ by considering $L^{\eta-2} \chi
\tilde{u}_h(\beta)^{-2}$. Such a combination is shown in Fig.~\ref{chi-HT}
(lower panel): all points fall on top of each other, confirming the validity
of the FSS Ansatz. Moreover, as expected, we find that $\tilde{C}(0)$ is
finite and $\tilde{C}(\xi/L)$ is approximately constant for $\xi/L \lesssim
0.15$, two properties which are not obvious from the upper panel of
Fig.~\ref{chi-HT}.  These conclusions are consistent with the FSS results for
$\chi(2L,\beta)/\chi(L,\beta)$ (in this quantity the analytic function
$\bar{u}_h(\beta)$ cancels out) reported in Refs.~\onlinecite{PC-99,Jorg-06},
which show that this ratio has a tiny dependence on $\xi/L$ up to
$\xi/L\approx 0.15$ (for $\xi/L = 0.15$ we have $\chi(2L,\beta)/\chi(L,\beta)
\approx 1.02$).  Note that the curve in the lower panel of Fig.~\ref{chi-HT}
(which corresponds to the dashed line in the upper panel) is the limiting
curve of the points that appear in the upper panel as $L\to \infty$. Since the
rate of convergence is very slow (at fixed $\xi/L$ data converge as
$L^{-1/\nu}$), it is clear that such an asymptotic behavior can only be
observed on enormously large lattices.  Thus, in order to estimate $\eta$, it
is crucial to take the function $\bar{u}_h(\beta)$ into account.

The function $\tilde{C}(x)$ is universal, apart from a model-dependent 
multiplicative constant. We write it as 
\begin{equation}
\tilde{C}(x) = b \Gamma(x) \qquad\qquad \Gamma(0) = 1,
\end{equation}
where $\Gamma(x)$ is universal. A fit of the data reported in 
Fig.~\ref{chi-HT} gives 
\begin{eqnarray}
&& \Gamma(x) = 
1  + 5.9622 y - 104.4625 y^2 
 + 1516.2443 y^3 \nonumber  \\
&& \qquad \quad - 12260.6638 y^4 + 50105.6104 y^5 
 - 80471.2150 y^6, \nonumber \\
&& y \equiv  \exp(-1/x),
\label{eqGamma}
\end{eqnarray}
and $b \approx 1.9395$.
The expression (\ref{eqGamma})
is valid for $x\equiv \xi/L\lesssim R^*_\xi \approx 0.645$.

The same analyses can be repeated for the $\pm J$ model at $p=0.7$
and the BDBIM at $p_b = 0.45$. Here we only present results 
corresponding to fits to the Ansatz (\ref{fita-chi}).
Our data are not precise enough to allow us to perform fits 
which include the nonanalytic scaling corrections. The results are 
consistent with the estimate $\eta = -0.375(10)$. 
For the $\pm J$ model at $p=0.7$,  we obtain 
$\eta = -0.366(2)$ ($\beta_{\rm min} = 0.59$) 
and $\eta = -0.366(3)$ ($\beta_{\rm min} = 0.68$) for $L_{\rm min} = 7$,
and 
$\eta = -0.368(2)$ ($\beta_{\rm min} = 0.59$)
and $\eta = -0.366(3)$ ($\beta_{\rm min} = 0.68$) for $L_{\rm min} = 9$.
For the BDBIM, we obtain 
$\eta = -0.361(2)$ and $-0.359(3)$ for $L_{\rm min} = 7$ and 9 
and any $\beta_{\rm min}$ in the range $[0.82,1.12]$.

\begin{figure}[tb]
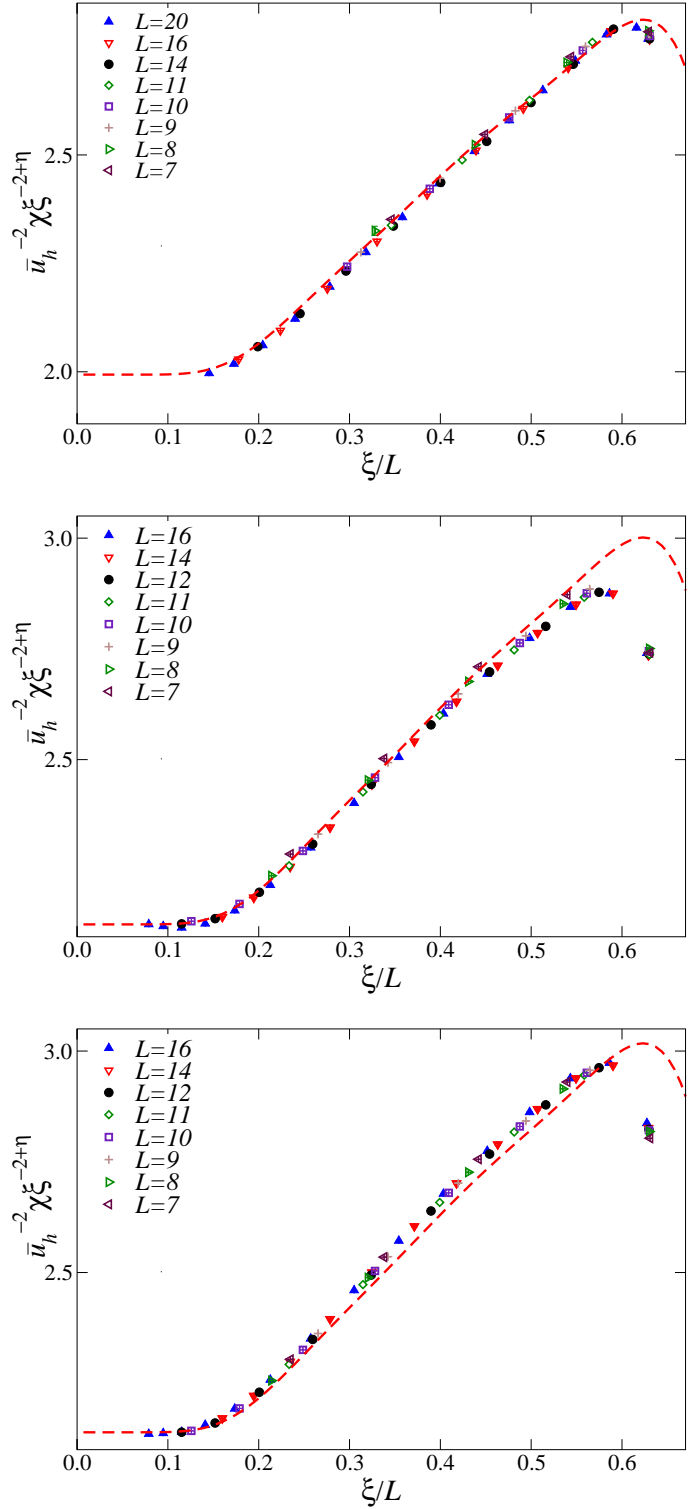

\centerline{\psfig{width=9truecm,angle=0,file=fig25a.eps}}
\vspace{4mm}
\centerline{\psfig{width=9truecm,angle=0,file=fig25b.eps}}
\vspace{4mm}
\centerline{\psfig{width=9truecm,angle=0,file=fig25c.eps}}
\caption{
  (Color online) 
   Plot of $\chi \xi^{\eta - 2} \tilde{u}_h^{-2}$ 
   vs $\xi/L$ for the $\pm J$ model at $p=0.7$ with $\eta = -0.375$ 
   (upper panel) and 
   for the BDBIM at $p_b = 0.45$ with $\eta = -0.375$
   (middle panel) and $\eta = - 0.360$ (lower panel).
   In each panel we report the curve $b \Gamma(\xi/L)$, 
   where $\Gamma(\xi/L)$ is defined in Eq.~(\ref{eqGamma})
   and $b$ is a model-dependent constant.
   We use $b = 1.99346$, 2.12794, 2.13923, in the upper, middle, and lower
   panel, respectively. They are determined by requiring a perfect fit 
   for $\xi/L \le 0.2$. 
  }
\label{FSSchi-altrimodelli}
\end{figure}

As a further check of universality we determine the scaling behavior of the
combination $\chi \xi^{\eta - 2} \tilde{u}_h^{-2}$.  In
Fig.~\ref{FSSchi-altrimodelli} we report this quantity for the $\pm J$ model
at $p=0.7$ (upper panel) and the BDBIM at $p_b = 0.45$ (middle panel), using
in both cases $\eta = -0.375$ and our best estimate of $\tilde{u}_h$. As
expected all points fall onto a single curve. If universality holds, these
curves should be parametrized as $b \Gamma(x)$, where $\Gamma(x)$ is given in
Eq.~(\ref{eqGamma}) and $b$ is a model-dependent constant.  In the case of the
$\pm J$ model (upper panel of Fig.~\ref{FSSchi-altrimodelli}) we observe a
very good agreement, while for the BDBIM at $p_b = 0.45$ (middle panel) some
discrepancies occur for $\xi/L\gtrsim 0.4$.  There are two reasons for them.
First, the function $\tilde{u}_h$ is not precisely known for $\beta \approx
\beta_c$: in this range of values of $\beta$ it varies slightly (10\%) with
$\beta_{\rm min}$ and $L_{\rm min}$.  Second, the plot depends on $\eta$. If
we use $\eta = -0.360$, which is the value obtained in the fits for the
diluted model which provide $\tilde{u}_h$, we obtain the lower panel of
Fig.~\ref{FSSchi-altrimodelli}.  Discrepancies are now significantly reduced.

It is worth noting that the functions $\bar{u}_h$ are approximately the same
in the three models we study, if one considers them as a function of the
reduced temperature $t \equiv 1-\beta/\beta_c$.  For $0\lesssim t
\lesssim 0.4$---this is the interval of $t$ which is probed by our
simulations---the ratio $\bar{u}_{h,\rm model 1}(t)/\bar{u}_{h,\rm model
  2}(t)$ is constant, within our precision, for any pair of models.  This
result is somewhat unexpected within RG theory, because these functions are
not universal.

Finally, let us comment on the FSS approach of Ref.~\onlinecite{CEFPS-95},
applied to spin-glass systems in Refs.~\onlinecite{PC-99,Jorg-06}. 
In this approach one considers the ratio $\chi(2L,\beta)/\chi(L,\beta)$.
This choice has a significant advantage: the scaling-field function 
$\bar{u}_h(\beta)$ cancels out, so that the leading scaling corrections
are the nonanalytic ones. As we have shown here, they are quite small, 
so that very good scaling is observed and reliable infinite-volume 
estimates are obtained. Analytic scaling corrections come in again when 
considering the critical limit of the infinite-volume results 
$\chi_\infty(\beta)$. Indeed, since $\Delta \equiv \omega \nu \approx 2.45$,
for $\beta\to\beta_c$ the analytic corrections dominate: 
\begin{equation}
\chi_\infty(\beta) = (\beta_c - \beta)^{-\gamma} \, 
  [b_0 + b_1 (\beta_c - \beta) + b_2 (\beta_c - \beta)^2 + 
   b_\Delta (\beta_c - \beta)^\Delta + \cdots ].
\end{equation}

\subsection{Extended-scaling scheme} \label{secVIII.C}

\begin{figure}[tb]
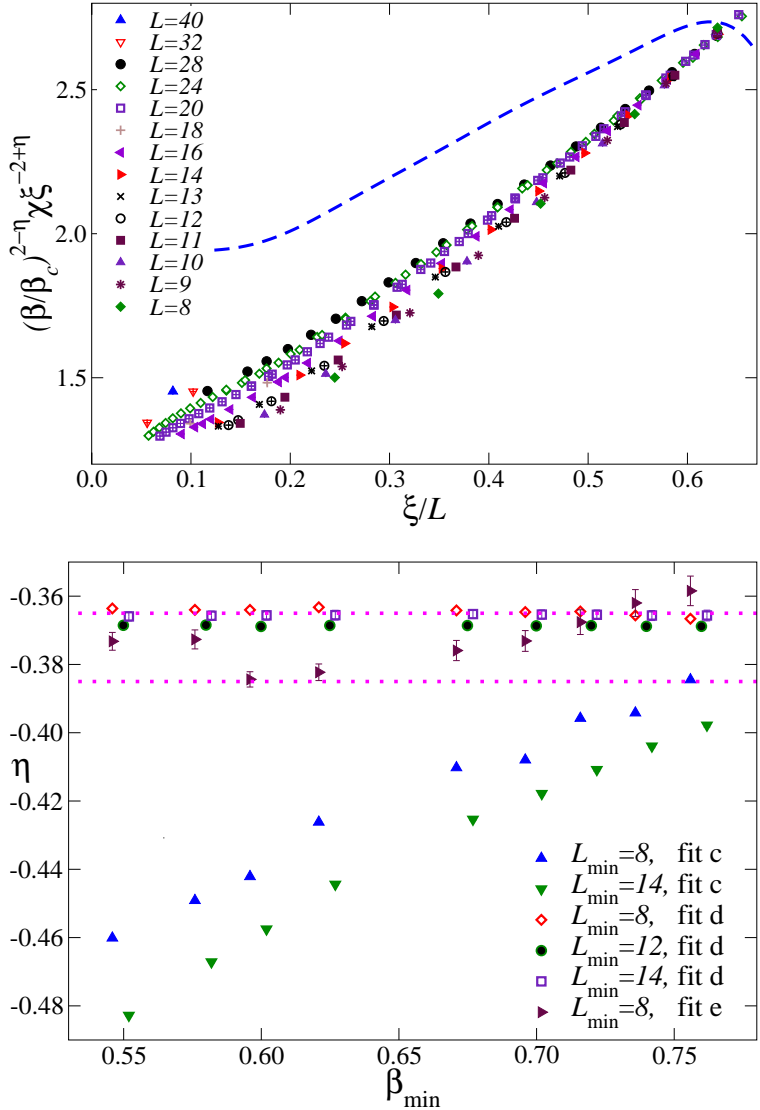

\centerline{\psfig{width=10truecm,angle=0,file=fig26a.eps}}
\vspace{4mm}
\centerline{\psfig{width=10truecm,angle=0,file=fig26b.eps}}
\vspace{2mm}
\caption{
  (Color online) Extended-scaling results for the 
  $\pm J$ model at $p = 0.5$. In the upper panel we report 
  $\beta^{2 - \eta} \chi \xi^{\eta - 2}$ vs $\xi/L$. 
  The dashed line is the universal curve $\tilde{C}(\xi/L)$, as 
  estimated in fits of $\chi$ to Eq.~(\ref{fita-chi}).
  In the lower panel we report estimates of $\eta$ obtained in three different
  fits for several values
  of $L_{\rm min}$ and $\beta_{\rm min}$. Fit c uses the 
  Ansatz (\ref{fitc-chi}), fit d and e are defined in text, 
  below Eq.~(\ref{fitc-chi}).
  The lines correspond to the estimate $\eta = -0.375(10)$.
  }
\label{etaCPB-HT}
\end{figure}

In this Section we consider the extended-scaling scheme introduced in
Ref.~\onlinecite{CHT-06}.  It consists in a particular choice of scaling
variables, which, according to Ref.~\onlinecite{CHT-06}, should somehow
decrease scaling corrections and thus allow a faster convergence to the
critical limit.  Let us consider first $\xi/L$. 
In this scheme 
the appropriate fit Ansatz is
\begin{equation}
{\xi(\beta,L)\over L} = 
   P_n(x)^{-\nu/n}, 
  \qquad \qquad x \equiv (\beta_c^2 - \beta^2) (L/\beta)^{1/\nu},
\label{fitCPB}
\end{equation}
where $P_n(x)$ is a polynomial of degree $n$.
The results for the $\pm J$ model at $p = 0.5$ 
are reported in Fig.~\ref{nu-betac-HT} and should be compared with
those obtained by fits to Eq.~(\ref{fita}), which neglect any scaling
corrections (fit a).  They are substantially equivalent and no improvement is
observed. This can be explained by noting that, for the present values of
$\beta_c$ and $\nu$, we have
\begin{equation}
(\beta_c^2 - \beta^2) \beta^{-1/\nu} = 
     1.88 (\beta_c - \beta) [ 1 - 0.10 (\beta_c - \beta) + \ldots ].
\end{equation}
Thus, fit (\ref{fitCPB}) is essentially equivalent to a fit with analytic
corrections (fit b) with $b = -0.10$. Such a value of $b$ is small---hence,
this change of the scaling variable does not have much influence on the final
results---and is close to what we obtain numerically, though not fully
consistent (we predict $0\lesssim b \lesssim 0.3$). 
We also tried fits with $x \equiv (\beta_c^2 - \beta^2)
L^{1/\nu}$, which should be the natural variable in spin-glass systems,
given the symmetry under $\beta\to -\beta$.\cite{DCA-04} 
These fits are significantly worse than the previous ones for
$\beta \lesssim 0.70$.  For larger values, no significant differences are
observed.  These results can be understood by noting 
$(\beta_c^2 - \beta^2) = 1.80
(\beta_c - \beta) [ 1 - 0.55 (\beta_c - \beta) + \ldots ]$. Thus, this choice
of scaling variable corresponds to assuming $b = -0.55$ in
Eq.~(\ref{expan-ut}), which is significantly larger than what we find
numerically. Therefore, if we use $(\beta_c^2 - \beta^2)$ as 
approximate thermal scaling field, the analytic corrections---in this case
they are proportional to $(\beta_c^2 - \beta^2)^2$---are more important
than in the case in which $u_t$ is simply approximated with 
$\beta_c - \beta$. 

The extended-scaling scheme can also be applied to the analysis of the 
susceptibility.  It amounts to consider 
the scaling Ansatz\cite{CHT-06}
\begin{equation}
\chi(\beta,L) = \beta^{\eta-2} \xi^{2-\eta} C(\xi/L).
\label{chiFSS-CPB}
\end{equation}
In Fig.~\ref{etaCPB-HT} we show $\beta^{2 - \eta} \chi \xi^{\eta -2}$
versus $\xi/L$ for the $\pm J$ model at $p=0.5$. 
Scaling is better than that observed in the upper panel of 
Fig~\ref{chi-HT}: the scatter of the data points
is significantly reduced, indicating that $\beta^{\eta-2}$ approximates 
the scaling-field term $\bar{u}_h^2$ better than a constant.
However, the rescaled data are still
far from the asymptotic curve (the dashed line)
determined numerically above, indicating that
the residual analytic scaling corrections are also in this case not 
negligible.
This is better understood, by comparing the function $\tilde{u}_h$, as 
determined in the fits, with the approximation $(\beta/\beta_c)^{\eta/2-1}$,
which follows from Eq.~(\ref{chiFSS-CPB}). As can be seen from 
Fig.~\ref{uh-fun}, the approximate expression proposed in 
Ref.~\onlinecite{CHT-06} has the correct qualitative shape, but differs 
significantly from the quantitative point of view. For these reasons,
we do not expect the scaling Ansatz (\ref{chiFSS-CPB}) to be particularly
useful to estimate $\eta$ from our data. 

To understand the role of the residual analytic scaling corrections on the
determinations of $\eta$, we fit the data for the $\pm J$ model at $p=0.5$ to
the scaling Ansatz (fit c)
\begin{equation}
\ln {\chi \beta^2\over \xi^2} = - \eta \ln {\xi\over \beta} + 
   P_n(\xi/L),
\label{fitc-chi}
\end{equation}
where $P_n(x)$ is a polynomial of degree $n$. The results are 
reported in Fig.~\ref{etaCPB-HT}. They vary strongly with 
$\beta_{\rm min}$ and $L_{\rm min}$, indicating that scaling corrections
are sizable and not negligible. As a test we have also repeated 
the fits to Eqs.~(\ref{fita-chi}) and (\ref{fitb-chi}) 
replacing $\ln {\chi/\xi^2}$ with $\ln {\chi \beta^2/ \xi^2}$ and 
$\ln \xi$ with $\ln \xi/\beta$ 
(we call fit d and fit e the fits corresponding to
Eqs.~(\ref{fita-chi}) and (\ref{fitb-chi}), respectively).
As expected, the results are identical to those obtained in fit b and 
fit c, respectively. Indeed, 
the fits only differ in the parametrization of the 
analytic function $\bar{u}_h$. Note that the extended-scaling approximation
for $\bar{u}_h$ is not analytic in $\beta$, since $\beta = 0$ is 
a branching point. This is, however, irrelevant in practice,
since we are looking for approximations of $\bar{u}_h$ in the interval
$0.6\lesssim \beta \lesssim 0.9$, which is quite far from $\beta = 0$.

\section{The zero-momentum quartic couplings $G_4$ and $G_{22}$ in
the high-temperature phase}
\label{HTTL}

\begin{figure}[tb]
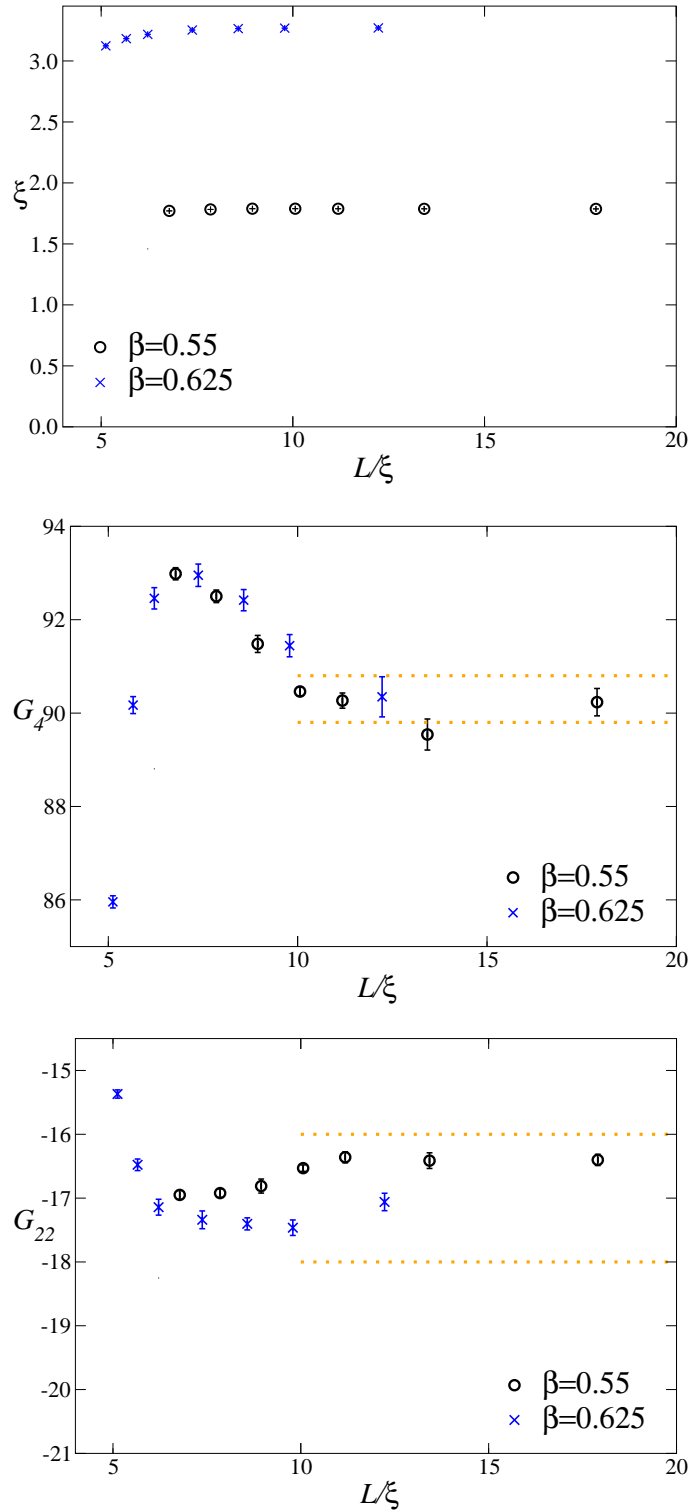

\centerline{\psfig{width=9truecm,angle=0,file=fig27a.eps}}
\vspace{4mm}
\centerline{\psfig{width=9truecm,angle=0,file=fig27b.eps}}
\vspace{4mm}
\centerline{\psfig{width=9truecm,angle=0,file=fig27c.eps}}
\vspace{2mm}
\caption{
  (Color online) Estimates of $\xi(L,\beta)$ (top), 
  $G_4(L,\beta)$ (middle), and $G_{22}(L,\beta)$ (bottom) versus 
  $L/\xi$ for 
  $\beta = 0.55$ and $\beta = 0.625$, corresponding to 
  the infinite-volume correlation lengths $\xi_\infty \approx 1.79$ 
  and $\xi_\infty \approx 3.27$. Results for the $\pm J$
  model at $p=0.5$. The dotted lines correspond to the 
  estimates $G_4^* = 90.3(5)$ (middle panel), $G_{22}^* = -17(1)$ (lower panel).
 }
\label{g4g22}
\end{figure}

In this section we consider 
the zero-momentum quartic couplings $G_4$ and $G_{22}$
defined in Eq.~(\ref{g4def}) and (\ref{g22def}) and estimate 
their infinite-volume critical value defined by
\begin{equation}
   G^*_\# = \lim_{\beta\to \beta_c^+} \lim_{L\to\infty} G_\#(L,\beta).
\end{equation}
We consider the $\pm J$ Ising model at $p=0.5$ and use the estimates of the
quartic couplings for $\beta=0.55$ and $L=12$, 14, 16, 18, and $32$, and for
$\beta=0.625$ and $L=16$, 18, 28, 32, and $40$.  We combine results obtained
in the random-exchange runs that we performed for our FSS study around
$\beta_c$ and results obtained in standard MC simulations for these two values
of $\beta$.  \cite{footnoteHT} First, we investigate the infinite-volume
limit. The correlation length converges rapidly (see Fig.~\ref{g4g22}). For
instance, for $\beta = 0.55$ we obtain $\xi=1.7888(3),\,1.7885(5),\,
1.7872(3)$ for $L=18,\,20,\,32$, while for $\beta = 0.625$ we have
$\xi=3.2694(12), 3.2709(13)$ for $L=32$ and $L=40$: for $L/\xi\gtrsim 10$ the
results vary by less than 0.1\%, indicating that the difference from their
thermodynamic limit is within 0.1\%.  Thus, we can take as infinite-volume
estimates those obtained on the largest lattices.  The quartic couplings show
larger finite-size corrections.  As shown in Figs.~\ref{g4g22} the
infinite-volume limit is approximately reached for $L/\xi\gtrsim 12$, within
our statistical precision.  Indeed at $\beta=0.625$ we find
$G_4=90.46(9),\,90.27(16),\,90.23(29)$ and
$G_{22}=-16.53(5),\,-16.36(9),\,-16.40(9)$ for $L=18,\,20,\,32$, respectively.
We can thus take the estimates on the largest lattices as infinite-volume
estimates.

In the critical limit we expect
\begin{equation}
G_\#(L=\infty,\beta) = G_\#^* + c_\# \xi^{-\omega},
\label{approtc}
\end{equation}
with $\omega=1.0(1)$.
The comparison of the results at $\beta=0.55$ and $\beta=0.625$ leads
us to the estimates 
\begin{equation}
G_4^* = 90.3(5),\qquad G_{22}^* = -17(1),
\label{g4g22est}
\end{equation}
where the error takes also into account the effects of the $O(\xi^{-\omega})$
scaling corrections, which is roughly estimated from the difference of the
infinite-volume results at the two values of $\beta$.  These results can be
compared with those obtained in Ref.~\onlinecite{PPV-06} for the
random-anisotropy Heisenberg model in the limit of infinite anisotropy: $G_4^*
= 88(8)$ and $G_{22}^* = -11(4)$.  There is a substantial agreement, which
confirms that the critical behavior of the random-anisotropy Heisenberg model
for infinite anisotropy belongs to the Ising spin-glass universality
class.\cite{PPV-06,LLMPV-07}

\section{Conclusions}
\label{conclusions}

In this paper we discuss the critical behavior of three-dimensional
Ising spin-glass systems with the purpose of verifying 
universality, clarifying the role of scaling corrections, and 
determining the critical exponents. More precisely,
our results can be summarized as follows.
\begin{itemize}
\item[i)] By using the RG we derive the behavior for 
$L\to \infty$ and $\beta\to \beta_c$ of several quantities which are 
routinely measured in MC simulations. In particular,
we show that the analytic dependence
of the scaling fields on the model parameters may give rise to 
corrections which behave as $L^{-1/\nu}\sim L^{-0.4}$. If they are 
neglected, FSS analyses give inconsistent results. These corrections
have been overlooked in previous FSS studies. Note that the general 
expressions we obtain are relevant also in other glassy systems, in which 
$\nu$ is typically large.
\item[ii)] We determine the leading nonanalytic correction-to-scaling
exponent $\omega$. We obtain $\omega = 1.0(1)$. Note that in 
Ising spin-glass systems nonanalytic scaling corrections decay faster than 
in the Ising model, in which $\omega\approx 0.8$, see Ref.~\onlinecite{PV-02}.
The exponent $\omega$ is also significantly larger than that at the PF 
transition, which occurs for small frustration: 
$\omega = 0.29(2)$.\cite{HPV-07}
\item[iii)] We accurately verify universality. A careful analysis of 
the quartic cumulants $U_4$ and $U_{22}$ at fixed $R_\xi = 0.63$
shows that their limit for $L\to\infty$ is independent of the model and 
of the disorder distribution. 
The results obtained in the different models differ at most by
approximately one per mille in the case of $\bar{U}_4$ and one per cent
for $\bar{U}_{22}$. They support the 
existence of a unique Ising spin-glass universality class.
Universality is also supported by the FSS analyses of $\xi$ and 
$\chi$ in the high-temperature phase. We verify that 
the FSS curves for these two quantities
are independent of the model.
\item[iv)] We determine the critical exponents. For this purpose
we perform analyses at the critical point and analyses which take into 
account all high-temperature data. Results are consistent, once 
the analytic and the nonanalytic corrections are taken into account. 
Moreover, they do not depend on the model and disorder distribution.
Again, this supports the universality of the paramagnetic-glassy transition.
We obtain 
\begin{equation}
\nu = 2.45(15), \qquad\qquad \eta = -0.375(10).
\label{crit-exp-final}
\end{equation}
Using scaling and
hyperscaling relations, we find $\beta=\nu(1+\eta)/2=0.77(5)$,
$\gamma=(2-\eta)\nu=5.8(4)$, and $\alpha=2-3\nu=-5.4(5)$.  
\end{itemize}
Our estimates of the critical exponents can be compared with those
reported in the literature. Earlier estimates of $\nu$ are reported in 
the introduction and in Table I of Ref.~\onlinecite{KKY-06}.
Some of the most recent ones are close to our final estimate. For the 
exponent $\eta$, we quote here the most recent results: $\eta =
-0.395(17)$, $-0.37(5)$,\cite{KKY-06} 
$\eta = -0.40(4)$,\cite{CHT-06} and 
$\eta = -0.349(18)$.\cite{Jorg-06} They are all in substantial agreement 
with our result, which is however significantly more precise. 
We also mention the estimate
$\beta=0.52(9)$, obtained in Ref.~\onlinecite{PRT-06} by an out-of-equilibrium
simulation. 

The estimates (\ref{crit-exp-final})
slightly differ from, and have larger errors than, those obtained in
Ref.~\onlinecite{HPV-08}: $\nu=2.53(8)$ and $\eta=-0.384(9)$. 
There are two reasons for that. First, we have significantly extended the runs
for $L=20,24$ for the $\pm J$ model at $p=0.5$. The new results have slightly
shifted the estimates of $\beta_c$, $R^*_\xi$, and of the critical 
exponents. Second, we have been more conservative. With our present
error bars, the estimates are fully consistent with the results of all analyses
for the three models we considered.

We also analyzed our data by using the 
extended-scaling scheme proposed in Ref.~\onlinecite{CHT-06}. 
This approach might partly take into account the scaling
corrections arising from the analytic dependence of the scaling fields on the
reduced temperature but it neglects the nonanalytic corrections arising
from the irrelevant operators. In some cases, for instance for the overlap 
susceptibility, 
this scheme shows an apparent improvement of the scaling behavior 
with respect to the naive approach in which the analytic corrections
are simply neglected. However, 
the approximate expressions which follow from the extended-scaling scheme
are not sufficiently precise for a high-precision study of the 
critical behavior: if one aims at accurate estimates, it is 
necessary to determine the corrections directly from the data.

\appendix

\section{Finite-size behavior of the phenomenological couplings}
\label{proofsca}

We now provide a detailed proof of Eq.~(\ref{Rexp_1}) for the phenomenological
couplings $U_4$, $U_{22}$, and $R_\xi\equiv \xi/L$. First, we present a simple
physical argument that clarifies the origin of the function $\bar{u}_h(t)$
introduced in Eq.~(\ref{uhsca}), and then, a more formal argument based on the
Wegner expansion.\cite{Wegner-76}

A short and physical proof goes as follows.  First, note that $\chi$ is not
RG invariant and, in particular, it is not invariant under field
redefinitions. Given a theory with fields $\phi$, define
\begin{equation}
\chi_\phi = \int d^3x [{\langle \phi(x) \phi(0)\rangle}].
\end{equation}
If we change variables $\phi = Z \psi$, we obtain 
\begin{equation}
\chi_\phi = Z^2 \int d^3x [{\langle \psi(x) \psi(0)\rangle}] = 
       Z^2 \chi_\psi.
\end{equation}
Thus, under a field redefinition $\chi \to Z^2 \chi$. The function 
$\bar{u}_h(t)$ is exactly the field redefinition that relates the 
bare lattice fields to the renormalized fields. Hence, we expect 
at criticality
\begin{equation}
   \chi = \bar{u}_h^2 L^{2 - \eta}.
\label{chibaru}
\end{equation}
A $t$ dependent prefactor is absent in the phenomenological couplings,
because 
these quantities are RG invariant and in particular, they do not depend on 
the normalization of the fields. This is quite obvious from their definitions.
In the continuum limit they can be written as 
\begin{eqnarray}
&&U_4 \equiv {\int dx_1 dx_2 dx_3 dx_4 
   [\langle \phi(x_1) \phi(x_2) \phi(x_3) \phi(x_4)\rangle] \over 
   \left[\int d x_1 dx_2 {\langle \phi(x_1) \phi(x_2) \rangle}
   \right]^2  },\\
&&U_{22} \equiv {\int d x_1 dx_2 dx_3 dx_4 
   [{\langle \phi(x_1) \phi(x_2)\rangle
         \langle \phi(x_3) \phi(x_4)\rangle}] \over 
   \left[\int d x_1 dx_2 {\langle \phi(x_1) \phi(x_2) \rangle}
   \right]^2 } \, - 1,\\
&&\xi^2 \equiv {1\over6} 
     {\int d x_1 dx_2 (x_1 - x_2)^2 
      [\langle \phi(x_1) \phi(x_2)\rangle] \over 
     \int d x_1 dx_2 
      {\langle \phi(x_1) \phi(x_2)\rangle} }.
\end{eqnarray}
The formula we report for $\xi$ correspond to Eq.~(\ref{xidefffxy}) for 
$L\to \infty$, disregarding corrections of order $L^{-2}$.
Since the number of fields in the denominator is equal to the number of fields
in the numerator in all expressions, each quantity is invariant under 
field redefinitions.

This discussion clarifies the origin of the 
$t$-dependent prefactor in Eq.~(\ref{chibaru})
and is at the basis of the statement, which is 
well accepted in the literature, that $U_4(T_c)$, $U_{22}(T_c)$,
and $\xi(T_c)/L$ approach {\em universal} constants as $L\to\infty$. 
The presence of an analytic prefactor would violate universality.
Indeed, imagine that a phenomenological coupling $R$ behaves as 
\begin{equation}
R(\beta,L) = \bar{u}_h(t)^{\rm some\, power} g(u_t L^{1/\nu}) + 
   \hbox{\rm scaling corrections}
\end{equation}
close to the critical point.
Since $\bar{u}_h(t)$ is model dependent, at the critical point $R(\beta_c,L)$
would not be universal for $L\to\infty$.
Hence, on purely physical grounds, there must be no $t$-dependent prefactor
in any phenomenological coupling, hence no analytic corrections.

This short discussion shows that 
$R_\xi$ and $\chi/L^{2-\eta}$ are conceptually two very different objects. 
The first quantity is RG invariant and shows a universal
FSS behavior. The second quantity is not RG invariant
and, for instance, $\chi(T_c)/L^{2-\eta}$ converges to a 
{\em model-dependent} constant as $L\to \infty$.

Eq.~(\ref{Rexp_1}) can also be derived from the usual Wegner's scaling expression 
for the free energy. Let us first consider $U_4$.
Using Eq.~(\ref{FscalL}), we find
\begin{eqnarray}
&&[\mu_4] = \left. L^d {\partial^4 {\cal F}\over \partial h^4} \right|_{h=0}= 
   L^{4y_h} \bar{u}_h^4 f^{(4)}(0,u_t L^{y_t}) + \cdots  \\
&&[\mu_2] = \left. L^d {\partial^2 {\cal F}\over \partial h^2} \right|_{h=0}=
    L^{2y_h} \bar{u}_h^2 f^{(2)}(0,u_t L^{y_t}) + \cdots   
\end{eqnarray}
[the dots correspond to nonanalytic scaling corrections and 
bulk contributions, and the derivatives
refer to the first variable appearing in the scaling function $f(x,y)$] so that 
\begin{equation}
U_4 = {f^{(4)}(0,u_t L^{y_t})\over f^{(2)}(0,u_t L^{y_t})^2} + \cdots ,
\end{equation}
which proves Eq.~(\ref{Rexp_1}).

To discuss $U_{22}$ one should generalize Wegner's scaling expression
(see Sec.~3.1 of Ref.~\onlinecite{HPPV-07} for a detailed discussion). 
Define $Z(\beta,h,L)$ as the partition function of two systems at
inverse temperature $\beta$ defined on a lattice of size $L^3$ coupled 
by an interaction  
\begin{equation}
h \sum_x \sigma_{1x} \sigma_{2x} .
\end{equation} 
Then, consider
\begin{equation}
{\cal F}(\beta,h_1,h_2,L) = 
   L^{-d} \left[\ln Z(\beta,h_1,L) \ln Z(\beta,h_2,L) \right].
\end{equation}
A scaling Ansatz like Eq.~(\ref{FscalL01})
allows one to obtain an expression analogous to that obtained for $U_4$
and to prove Eq.~(\ref{Rexp_1})  for $U_{22}$.

In order to determine the scaling behavior of $R_\xi$ we  
consider a momentum-dependent magnetic field. The argument goes as follows:
Define $Z(\beta,h,L,p)$ as the partition function of two systems at
inverse temperature $\beta$ defined on a lattice of size $L^3$ coupled 
by an interaction $h \sum_x \sigma_{1x} \sigma_{2x} \cos (p\cdot x)$. 
Then, consider the corresponding disorder-averaged free-energy density
\begin{equation}
{\cal F}(\beta,h,L,p) = L^{-d}
   \left[\ln Z(\beta,h,L,p)\right].
\end{equation}
Under RG transformations $L\to \lambda L$, momenta scale as
$p \to p/\lambda$, so that the singular part of the free-energy density 
scales as 
\begin{equation}
{\cal F}_{\rm sing}(\beta,h,L,p) = L^{-d}
    f(pL,u_h(h,t,p) L^{y_h}, u_t(h,t,p) L^{y_t}),
\end{equation}
where we have neglected the nonanalytic scaling corrections and now
the scaling fields depend also on $p$.
Taking derivatives with respect to $h$ and then setting $h = 0$,
we obtain for the 
two-point function [of course $u_h(h,t,-p) = u_h(h,t,p)$]
\begin{equation}
\widetilde{G}(p) = \bar{u}_h(t,p)^2 L^{2-\eta} 
       f^{(2)}(pL,0,u_t(0,t,p) L^{y_t}),
\end{equation}
where we write as before
\begin{equation}
u_h(h,t,p) = h \bar{u}_h(t,p) + O(h^3),
\end{equation}
and we have neglected subleading terms.
For $p\to 0$, because of the cubic symmetry of the lattice, we have 
\begin{eqnarray}
&&\bar{u}_h(t,p) = \bar{u}_h(t) + O(p^2),\\
&&u_t(0,t,p) = u_t(0,t) + O(p^2),
\end{eqnarray}
where $\bar{u}_h(t)$ and $u_t(0,t)$ are the usual (zero-momentum) scaling
fields.
Hence, for $p\to 0$, disregarding corrections of order $p^2$, we can express 
$\widetilde{G}(p)$ in terms of the scaling fields that appear for 
$p = 0$: 
\begin{equation}
\widetilde{G}(p) = \bar{u}_h(t)^2 L^{2-\eta}
       f^{(2)}(pL,0,u_t(0,t) L^{y_t}) + O(p^2) + \cdots
\end{equation}
In the definition (\ref{xidefffxy}) of the correlation length $\xi$
we should consider 
$p = q \sim {1/L}$. Thus, disregarding terms of order $L^{-2}$ we have
\begin{equation}
{\widetilde{G}(0) - \widetilde{G}(p) \over \widetilde{G}(p)} = 
   {f^{(2)}(0,0,u_t(0,t) L^{y_t}) 
    \over f^{(2)}((2\pi,0,0),0,u_t(0,t) L^{y_t})} -1
   = \Phi(u_t(0,t) L^{y_t}).
\end{equation}
Neglecting again corrections of order $1/L^2$, we have 
\begin{equation}
{\xi^2\over L^2} = {1\over 4\pi^2}  \Phi(u_t(0,t) L^{y_t}), 
\end{equation}
which proves Eq.~(\ref{Rexp_1}).
If we consider the corrections to scaling, this derivation shows that 
$R_\xi$ behaves essentially as $U_{22}$ and $U_4$. The only difference is 
the presence of corrections due to the momentum-dependence of the scaling
fields and to the specific definition of the correlation
length: they scale as $L^{-2}$, $L^{-4}$, $\ldots$, $L^{-\omega-2}, \ldots$
Since $\omega \approx 1$, in Eq.~(\ref{Rexp_1}) they represent 
additional subleading corrections and can thus be neglected. This allows us
to consider $R_\xi$, $U_{22}$, and $U_4$ on the same footing.

\section{Some technical details on the MC simulations}
\label{MCinfo}

In our MC simulations we implement the standard Metropolis algorithm with a
sequential update of the spins. We use a multispin \cite{multispin}
implementation, in which $n_{\rm bit}=64$ systems are simulated in parallel.
For each of them we use a different set of bonds $\{J_{xy}\}$.

For the random numbers we use the SIMD-oriented Fast Mersenne Twister (SFMT)
\cite{twister} generator.  In particular, we use the {\sl genrand$\_$res53()}
function that produces double-precision output.  Independent random numbers
are employed to generate the starting configurations for each disorder
realization and in the parallel-tempering updates.  In the latter case very
few random numbers are used and thus, it takes virtually no extra time to use
individual random numbers for each of the $n_{\rm bit}$ systems
which are simulated in parallel.
On the other hand, in order to
save CPU time, we use the same sequence of random numbers for the 
local Metropolis update of
any of the $n_{\rm bit}$ systems. Though this choice does not lead to wrong
estimates of the expectation values, it might create a statistical correlation
among the $n_{\rm bit}$ systems. However, since each of the $n_{\rm bit}$
systems correspond to a different set of bond couplings $J_{xy}$ we expect
this effect to be negligible. Nevertheless, in order to ensure a correct
estimate of the statistical error, in our jackknife analysis we put all
$n_{\rm bit}$ systems that use the same sequence of random numbers in the same
bin.  In order to compute overlap observables, we performed runs for two
systems with the same set $\{J_{xy}\}$ in parallel.  In our MC simulations a
single Metropolis update of a single spin takes about $1.2 \times 10^{-9}$
seconds on an Opteron CPU running at 2 GHz (this should be compared with the
speed of the dedicated computer Janus,\cite{Belletti-etal-08} the fastest
computer simulating discrete spin models, which takes $2\times 10^{-11}$
seconds to update an Ising spin).

To reduce autocorrelations we used the random-exchange
or parallel-tempering method.\cite{raex}  To this end, we divided the
interval $[\beta_{\rm min}, \beta_{\rm max}]$ into $N_{\beta}-1$ equal
intervals $\Delta \beta$.  The parameter $\beta_{\rm max}$ 
was chosen such that $\xi(\beta_{\rm max})/L \approx 0.63$ in most
cases; in the latest runs we considered larger values,
such that $\xi(\beta_{\rm max})/L \approx 0.66$.
The parameter $\beta_{\rm min}$ was chosen such that 
$\xi(\beta_{\rm min}) \ll L$. 
We computed the observables in the neighborhood of $\beta_{\rm max}$ by using
their second-order Taylor expansion around $\beta_{\rm max}$. The coefficients of
the expansion were estimated by measuring appropriate correlators with the
energy.

An elementary update unit consists in  $n_{\rm met}$ Metropolis
sweeps followed by a replica exchange of all pairs of systems at nearby
temperatures.  The different systems were 
sequentially visited, starting from those at $\beta_{\rm min}$ and
$\beta_{\rm min}+\Delta \beta$. As a candidate for the exchange, we considered
one of two replicas with equal probability.  The acceptance probability for the
exchange is $\mbox{min}[1,\exp(-\Delta \beta \Delta H)]$.
Since the measurement of the energy
in our implementation costs more CPU time than a Metropolis sweep, we chose
$n_{\rm met}\gg 1$ independent of $\beta$.  

The computation of disorder averages of products of thermal expectations
requires particular care.  Indeed, naive estimators show a bias which may
become larger than statistical errors.\cite{BFMMPR-98-b} To avoid the problem
we consider essentially bias-free estimators, defined following
Ref.~\onlinecite{HPPV-07}.  For this purpose we divide the measurement phase
of the run into 12 intervals. Between each pair of subsequent interval there
is a decorrelation phase.  In total, the run consists of the following phases:
$Eq$, $D_1$ $M_1$, $D_2$ $M_2$, ..., $D_{12}$ $M_{12}$.  After some tests, we
fixed the number of update steps for each of them.  The equilibration phase
$Eq$ corresponds to $20n_{\rm temp}$ elementary update units; the measurement
phases $M_i$ correspond to $n_{\rm temp}$ update units, while the length of
$D_i$ is $n_{\rm temp}$ for $i\not=7$ and $5 n_{\rm temp}$ for $i=7$.  Recall
that each elementary update unit corresponds to $n_{\rm met}$ Metropolis
sweeps of all systems and to one full tempering sweep.

The presence of different measurement phases allows us to define
bias-free quantities. To define $[\langle A\rangle \langle B\rangle]$
(for instance, this is relevant for the computation of $U_{22}$)
we average over the samples the quantity
\begin{eqnarray}
{1\over 2\times 6\times 6} 
   \sum_{i=1}^6 \sum_{j=7}^{12} [\mu(A)_i \mu(B)_j + \mu(A)_j \mu(B)_i],
\end{eqnarray}
where $\mu(A)_i$ is the average of estimates of $A$ obtained in the 
measurement phase $M_i$. Analogously, to compute 
$[\langle A\rangle \langle B\rangle \langle C\rangle ]$ 
(these correlators are necessary to compute the coefficients of the 
Taylor expansions around a given value of $\beta$), we average 
over the samples the quantity 
\begin{eqnarray}
{1\over 3!\times 4^3}
   \sum_{i=1}^4 \sum_{j=5}^{8} \sum_{k=9}^{12} 
   [\mu(A)_i \mu(B)_j \mu(C)_k + \hbox{5 permutations}].
\end{eqnarray}

\begin{table}
\squeezetable
\caption{\label{parameters05}
Summary of the parameters for the runs at $p=0.5$. $N_{\beta}$ is the number 
of $\beta$ values used in the parallel-tempering simulation. 
The rest of the notation is 
explained in the text. The CPU
time refers to a single core of a dual core Opteron CPU running at 2.4 GHz.
}
\label{tempeinfo}
\begin{ruledtabular}
\begin{tabular}{rrrrrllr}
\multicolumn{1}{c}{$L$}&
\multicolumn{1}{c}{samples/64}&
\multicolumn{1}{c}{$n_{\rm met}$}&
\multicolumn{1}{c}{$n_{\rm temp}$}&
\multicolumn{1}{c}{$N_\beta$}&
\multicolumn{1}{c}{$\beta_{\rm min}$}&
\multicolumn{1}{c}{$\beta_{\rm max}$}&
\multicolumn{1}{c}{CPU time(days)}\\
\colrule
 4  &  100000 &   5 & 40   &  5  & 0.58& 0.92 &   \\
 5  &  100000 &   5 & 50   &  5  & 0.58& 0.908 &    \\
 6  &  100000 &   5 & 50   &  5   & 0.58& 0.9018 & 1 \\
 7  &  119103 &   5 & 80   &  5   & 0.58& 0.899 &  3 \\
 8  &  100000 &   5 & 80   &  5   & 0.58& 0.8975 & 4 \\
 9  &  110850 &  10 & 100  &  8   & 0.55& 0.8962 & 27 \\
10  &  100681 &  10 & 150  &  8   & 0.55& 0.896 & 50  \\  
11  &  109779 &  10 & 300  & 10   & 0.54& 0.8955 & 183 \\
12  &  106812 &  10 & 400  & 10   & 0.54& 0.8955 & 308 \\
13  &   38282 &  10 & 600  & 10   & 0.54& 0.8955 & 210 \\
14  &   31600 &  50 & 200  & 10   & 0.62& 0.8955 & 361 \\
16  &   24331 &  10 & 1000 & 20   & 0.52& 0.895 & 830 \\
20  &    1542 &  20 & 2000 & 32   & 0.5125 & 0.895 & 658  \\
20  &    2291 &  50 & 1500 & 20   & 0.625 & 0.91 & 1146  \\
24  &     717 &  25 & 2500 & 32   & 0.5125 & 0.895 & 826  \\
24  &    1627 &  50 & 2000 & 20   & 0.625 & 0.91 & 1874 \\
28  &     285 &  60 & 2500 & 20   & 0.6575 & 0.895 & 782  \\
\end{tabular}
\end{ruledtabular}
\end{table}

\noindent
In order to check equilibration, and decorrelation for the bias correction, we
followed the suggestion of Ref.~\onlinecite{KKY-06}. We doubled the length of
the run until the estimates of all observables were consistent within 
error bars.  We performed this check
only for the observables at $\beta_{\rm max}$, because these are expected to
be the most difficult ones for equilibration and decorrelation.  Starting from
disordered configurations, we determined the number of update steps $n_{\rm
  half}$ that are needed to reach (averaged over samples) half of the
equilibrium value of the overlap susceptibility.  In total, the equilibration
consisted of at least $100 n_{\rm half}$ update steps. 
Using these methods to check
equilibration, we came up with the choices summarized in Table
\ref{parameters05}. The parameters are not highly tuned, since we had
the CPU time available on short notice.  The runs that were done later have
typically a larger $\beta_{\rm min}$ than those done earlier. 
The run for $L=28$ is
a bit at the edge of the criterion given above for equilibration. However,
given the rather small number of samples ($N_s = 18240$), 
we are quite confident that the estimates  are correct within the quoted 
error bars.



\begin{thebibliography}{99}

\bibitem{EA-75}
S. F. Edwards and P. W. Anderson,
J. Phys. F {\bf 5}, 965 (1975).


\bibitem{IATKST-86}
A. Ito, H. Aruga, E. Torikai, M. Kikuchi, Y. Syono, H. Takei, 
Phys. Rev. Lett. {\bf 57}, 483 (1986).

\bibitem{GSNLAI-91}
K. Gunnarsson, P. Svedlindh, P. Nordblad, L. Lundgren,
H. Aruga, and A. Ito, Phys. Rev. B {\bf 43}, 8199 (1991).

\bibitem{NN-07}
S. Nair and A. K. Nigam,
Phys. Rev. B {\bf 75}, 214415 (2007).


\bibitem{OM-85}
A. T. Ogielski and I. Morgenstern,
Phys. Rev. Lett. {\bf 54}, 928 (1985).

\bibitem{Ogielski-85}
A. T. Ogielski, Phys. Rev. B {\bf 32}, 7384 (1985).

\bibitem{McMillan-85}
W. L. McMillan, Phys. Rev. B {\bf 31}, 340 (1985).

\bibitem{BM-85}
A. J. Bray and M. A. Moore, Phys. Rev. B {\bf 31}, 631 (1985).

\bibitem{BY-85}
R. N. Bhatt and A. P. Young,
Phys. Rev. Lett. {\bf 54}, 924 (1985).

\bibitem{SC-86}
R. R. P. Singh and S. Chakravarty,
Phys. Rev. Lett. {\bf 57}, 245 (1986).

\bibitem{RZ-86}
J. D. Reger and A. Zippelius,
Phys. Rev. Lett. {\bf 57}, 3225 (1986).

\bibitem{BY-88}
R. N. Bhatt and A. P. Young,
Phys. Rev. B {\bf 37}, 5606 (1988).

\bibitem{MPR-94}
E. Marinari, G. Parisi, and F. Ritort,
J. Phys. A {\bf 27}, 2687 (1994).

\bibitem{KY-96}
 N. Kawashima and A. P. Young,  Phys. Rev. B {\bf 53}, R484 (1996).

\bibitem{BPC-96}
L. W. Bernardi, S. Prakash, and I. A. Campbell,
Phys. Rev. Lett. {\bf 77}, 2798 (1996).

\bibitem{IPR-96}
D. I\~niguez, G. Parisi, and J. J. Ruiz-Lorenzo,
J. Phys. A {\bf 29}, 4337 (1996).

\bibitem{BJ-98}
B. A. Berg and W. Janke,
Phys. Rev. Lett. {\bf 80}, 4771 (1998).

\bibitem{MPR-98}
E. Marinari, G. Parisi, and J. J. Ruiz-Lorenzo,
Phys. Rev. B {\bf 58}, 14852 (1998).

\bibitem{MC-99}
P. O. Mari and I. A. Campbell, 
Phys. Rev. E {\bf 59}, 2653 (1999).

\bibitem{PC-99}
M. Palassini and S. Caracciolo,
Phys. Rev. Lett. {\bf 82}, 5128 (1999).

\bibitem{BCFMPRTTUU-00}
H. G. Ballesteros,
A. Cruz, L. A.~Fern\'andez, V.~Mart{\'\i}n-Mayor,
J.~Pech, J. J.~Ruiz-Lorenzo, A.~Taranc\'on, P.~T\'ellez,
C. L.~Ullod, and C.~Ungil, 
Phys. Rev. B {\bf 62}, 14237 (2000).

\bibitem{MC-02}
P. O. Mari and I. A. Campbell,
Phys. Rev. B {\bf 65}, 184409 (2002).

\bibitem{NEY-03}
T. Nakamura, S.-i. Endoh, and T. Yamamoto,
J. Phys. A {\bf 36}, 10895 (2003).

\bibitem{DCA-04}
D. Daboul, I. Chang, and A. Aharony, Eur. Phys. J. B {\bf 41}, 231 (2004).

\bibitem{PC-05}
M. Pleimling and I. A. Campbell, Phys. Rev. B {\bf 72}, 184429 (2005).

\bibitem{PRT-06}
S. Perez Gaviro, J. J. Ruiz-Lorenzo, and A. Taranc\'on,
J. Phys. A {\bf 39}, 8567 (2006).

\bibitem{PPV-06}
F. Parisen Toldin, A. Pelissetto, and E. Vicari,
J. Stat. Mech.: Theory Exp. P06002 (2006).

\bibitem{Jorg-06}
T. J\"org, Phys. Rev. B {\bf 73}, 224431 (2006).

\bibitem{CHT-06}
I. A. Campbell, K. Hukushima, and H. Takayama,
Phys. Rev. Lett. {\bf 97}, 117202 (2006).

\bibitem{KKY-06}
H. G. Katzgraber, M. K\"orner, and A. P. Young, 
Phys. Rev. B {\bf 73}, 224432 (2006).

\bibitem{MNS-08}
J. Machta, C. M. Newman, and D. L. Stein,
J. Stat. Phys. {\bf 130}, 113 (2008); arXiv:0805.0794.

\bibitem{HPV-08}
M. Hasenbusch, A. Pelissetto, and E. Vicari,
J. Stat. Mech.: Theory Exp. L02001 (2008).

\bibitem{Belletti-etal-08}
F. Belletti, M. Cotallo, A. Cruz, L. A. Fern\'andez, A. Gordillo-Guerrero, 
M. Guidetti, A. Maiorano, F. Mantovani, E. Marinari, V. Mart\'\i n-Mayor, 
A. Mu\~noz Sudupe, D. Navarro, G. Parisi, S. Perez-Gaviro, J. J. Ruiz-Lorenzo, 
S. F. Schifano, D. Sciretti, A. Taranc\'on, R. Tripiccione, J. L. Velasco, 
D. Yllanes (the Janus Collaboration),
Comp. Phys. Comm. {\bf 178}, 208 (2008).

\bibitem{CL-77}
K. H. Chen and T. C. Lubensky,
Phys. Rev. B {\bf 16}, 2106 (1977).

\bibitem{LLMPV-07}
F. Liers, J. Lukic, E. Marinari, A. Pelissetto, and E. Vicari,
Phys. Rev. B {\bf 76}, 174423 (2007).

\bibitem{Wegner-76} 
F.~J.~Wegner, in
{\em Phase Transitions and Critical Phenomena},
edited by C.~Domb and M.~S.~Green 
(Academic Press, New York, 1976), Vol.\ 6.

\bibitem{CHT-07}
I.~A. Campbell, K. Hukushima, and H. Takayama,
Phys. Rev. B {\bf 76}, 134421 (2007).

\bibitem{Nishimori-81}
H. Nishimori, Prog. Theor. Phys. {\bf 66}, 1169 (1981).

\bibitem{GHDB-85}
A. Georges, D. Hansel, P. Le Doussal, and 
J. Bouchaud, J. Phys. (Paris) {\bf 46}, 1827 (1985).

\bibitem{LH-88}
P. Le Doussal and A. B. Harris,
Phys. Rev. Lett. {\bf 61}, 625 (1988);
Phys. Rev. B {\bf 40}, 9249 (1989).

\bibitem{Nishimori-book}
H. Nishimori, {\em Statistical Physics of Spin Glasses and 
Information Processing: An Introduction}\/
(Oxford University Press, Oxford, 2001).

\bibitem{HPPV-07-mgp}
M. Hasenbusch, F. Parisen Toldin, A. Pelissetto, and E. Vicari,
Phys. Rev. B {\bf 76}, 184202 (2007).

\bibitem{DB-03}
Y. Deng and H. W. J. Bl\"ote,
Phys. Rev. E  {\bf 68}, 036125 (2003).

\bibitem{HPPV-07-pmj}
M. Hasenbusch, F. Parisen Toldin, A. Pelissetto, and E. Vicari,
Phys. Rev. B {\bf 76}, 094402 (2007). 

\bibitem{CPRV-02}
M. Campostrini, A. Pelissetto, P. Rossi, and E. Vicari,
Phys. Rev. E {\bf 65}, 066127 (2002).

\bibitem{HPPV-07}
M. Hasenbusch, F. Parisen Toldin, A. Pelissetto, and E. Vicari,
J. Stat. Mech.: Theory Exp. P02016 (2007).

\bibitem{PV-02}
A. Pelissetto and E. Vicari,
Phys. Rep. {\bf 368}, 549 (2002).

\bibitem{Toulouse-80}
G. Toulouse, J. Physique Lettres {\bf 41}, 447 (1980).

\bibitem{KR-03}
N. Kawashima and H. Rieger,  in 
{\em Frustrated Spin Systems}, edited by H. T. Diep
(World Scientific, Singapore, 2004); cond-mat/0312432.

\bibitem{Nishimori-86}
H. Nishimori, J. Phys. Soc. Japan {\bf 55}, 3305 (1986).

\bibitem{Kitatani-92}
H. Kitatani, J. Phys. Soc. Japan {\bf 61}, 4049 (1992).

\bibitem{WHP-03}
C. Wang, J. Harrington, and J. Preskill,
Ann. Phys. {\bf 303}, 31 (2003).

\bibitem{AH-04}
C. Amoruso and A. K. Hartmann,
Phys. Rev. B {\bf 70}, 134425 (2004).

\bibitem{PHP-06}
M. Picco, A. Honecker, and P. Pujol,
J. Stat. Mech.: Theory Expt. P09006 (2006).

\bibitem{Hartmann-99}
A. K. Hartmann, Phys. Rev. B {\bf 59}, 3617 (1999).

\bibitem{Kitatani-94}
It can be shown rigorously that the Nishimori line never intersects the 
spin-glass phase, H. Kitatani, J. Phys. Soc. Japan 
{\bf 63}, 2070 (1994). Since we must also have $p_{FG}\le p^*$,
the mixed phase, if it exists, should be 
confined to the region below the Nishimori line and on the left of 
the line $p = p^*$ (see Fig.~\ref{phdiad3}). 

\bibitem{SK-75}
D. Sherrington and S. Kirkpatrick,
Phys. Rev. Lett. {\bf 35}, 1792 (1975).

\bibitem{CKR-05}
T. Castellani, F. Krzakala, and F. Ricci Tersenghi,
Eur. Phys. J. B {\bf 47}, 99 (2005).

\bibitem{KM-02}
F. Krzakala and O. C. Martin, 
Phys. Rev. Lett. {\bf 89}, 267202 (2002).

\bibitem{BM-08}
S. Boettcher and E. Marchetti,
Phys. Rev. B {\bf 77}, 100405(R) (2008).

\bibitem{BD-08}
S. Boettcher and J. Davidheiser,
Phys. Rev. B {\bf 77}, 214432 (2008).

\bibitem{LZ-98}
C. D. Lorenz and R. M. Ziff, Phys. Rev. E {\bf 57}, 230 (1998).

\bibitem{JR-08}
T. J\"org and F. Ricci-Tersenghi,
Phys. Rev. Lett. {\bf 100}, 177203 (2008).

\bibitem{Privman-90}
V.~Privman, in
{\em Finite Scaling and Numerical Simulations of
Statistical Systems}, edited by V.~Privman
(World Scientific, Singapore, 1990).

\bibitem{SS-00}
J. Salas and A. D. Sokal, J. Stat. Phys. {\bf 98}, 551 (2000).

\bibitem{CHPV-06}
M. Campostrini, M. Hasenbusch, A. Pelissetto, and E. Vicari,
Phys. Rev. B {\bf 74}, 144506 (2006).

\bibitem{Has-99}
M. Hasenbusch, J. Phys. A {\bf 32}, 4851 (1999).

\bibitem{raex} C. J. Geyer in {\em Computer Science and Statistics: Proc. of
    the 23rd Symposium on the Interface}, edited by E. M.~Keramidas \/
  (Interface Foundation, Fairfax Station, 1991), p. 156; K. Hukushima and
  K. Nemoto, J. Phys. Soc. Jpn. {\bf 65}, 1604 (1996); 
  for a review, see 
  D. J. Earl and M. W. Deem, Phys. Chem. Chem. Phys. {\bf 7}, 3910 (2005).

\bibitem{BFMMPR-98-b}
H.G. Ballesteros, L. A. Fern\'andez, V. Mart\'{\i}n-Mayor, A. Mu\~noz~Sudupe,
G. Parisi, and J. J. Ruiz-Lorenzo,
Nucl. Phys. B {\bf 512}, 681 (1998).

\bibitem{foot-fit}
In practice, we write $R(L,\beta) = R(L,\beta_{\rm run}) + a_R(L) 
(\beta - \beta_{\rm run}) + b_R(L) (\beta - \beta_{\rm run})^2$,
where $\beta_{\rm run}$ is the value of $\beta$ closest to $\beta_c$ 
we have considered in the simulation (typically 
$\beta_{\rm run} = \beta_{\rm max}$, see Sec. \ref{mcsim}) and 
$R(L,\beta_{\rm run})$, $a_R(L)$, and $b_R(L)$ are determined in the 
simulation. Then, $\beta_c$, $b$, and $R^*$ are determined by minimizing
the usual $\chi^2$ variable.

\bibitem{CMPV-03}
P. Calabrese, V. Mart\'{\i}n-Mayor, A. Pelissetto, and E. Vicari,
Phys. Rev. E {\bf 68}, 036136 (2003).

\bibitem{footnote-fita} In the fit we minimize 
$\sum_i \left[ \Xi_i - P_n(x_i)\right]^2/\sigma_i^2$,
where $\Xi = (\xi/L)^{-n/\nu}$, 
the sum is over all data points, 
$\sigma_i = (n/\nu) \Xi_i \Delta \xi_i/\xi_i$,
and $\Delta \xi_i$ is the error on $\xi_i$. Errors are obtained 
by using a jackknife procedure and take into account the 
statistical correlations among the data. At fixed $\nu$ and $\beta_c$ 
the fit is linear and the result is obtained by inverting 
the least-square matrix. This matrix is ill-conditioned
(the condition number is of order $10^{24}$-$10^{32}$, depending on 
$n$). This required the use of 128-digit arithmetics in the calculations.

\bibitem{CEFPS-95}
S. Caracciolo, R. Edwards, S. J. Ferreira, A. Pelissetto, and 
A. D. Sokal, Phys. Rev. Lett. {\bf 74}, 2969 (1995).

\bibitem{footnoteHT} The finite thermodynamic limit of $G_{22}$ indicates
  that self-averaging holds in the high-temperature phase. Therefore, as the
  lattices become larger at a given temperature, it becomes convenient to
  improve the statistical accuracy of the expectation values for a given 
  sample.
  To this end we simulate $n>2$ copies of the system with 
  the same couplings $J_{xy}$. We compute the overlap 
  from all $n (n-1)/2$ pairs.

\bibitem{HPV-07}
M. Hasenbusch, A. Pelissetto, and E. Vicari,
J. Stat. Mech.: Theory Exp. P11009 (2007).

\bibitem{multispin}
See, e.g., S. Wansleben, J.B. Zabolitzky, and C. Kalle, J. Stat. Phys. {\bf 37},
271 (1984); G. Bhanot, D. Duke, and R. Salvador, 
Phys. Rev. B {\bf 33} 7841 (1986).

\bibitem{twister}
The numerical program and a detailed description can be found at 
``http://www.math.sci.hiroshima-u.ac.jp/{\~{}}m-mat/MT/SFMT/index.html".

\end{thebibliography}
\end{document}